\documentclass[10pt]{article}

\usepackage{amssymb}
\usepackage{graphicx}
\usepackage{dcolumn}
\usepackage{bm}
\usepackage{amsfonts}
\usepackage{amsmath}
\usepackage{amsthm}

\begin{document}

\title{Can persistent Epstein-Barr virus infection induce Chronic Fatigue Syndrome as a Pavlov reflex of the immune response?}

\author{Elena Agliari \\ \small{Dipartimento di Fisica, Universit\`a degli Studi di
Parma, viale G.P. Usberti 7/A, 43100 Parma (Italy)} \\
\small{INFN, Gruppo di Parma (Italy)}  \and  Adriano Barra \\ \small{Dipartimento di Fisica, Sapienza Universit\`a di Roma, Piazzale A. Moro 2, 00185 Roma (Italy)} \\ \small{GNFM, Gruppo di Roma 1 (Italy)} \and Kristian Gervasi Vidal \\ \small{IMT Institute for Advanced Studies, Piazza S. Ponziano 6, 55100 Lucca (Italy)} \and Francesco Guerra \\ \small{Dipartimento di Fisica, Sapienza Universit\`a di Roma, Piazzale A. Moro 2, 00185 Roma (Italy)} \\ \small{INFN, Gruppo di Roma 1 (Italy)} }
%
%
%
%
%

\maketitle

\begin{abstract}
Chronic Fatigue Syndrome is a protracted illness condition (lasting even years)
appearing with strong flu symptoms and systemic defiances by the immune system.\\
Here, by means of statistical mechanics techniques, we study the most widely accepted picture for its genesis, namely a persistent acute mononucleosis
infection, and we show
how such infection may drive the immune system toward an out-of-equilibrium metastable state displaying chronic
activation of both humoral and cellular responses (a state of full inflammation without a direct
"causes-effect" reason).\\
By exploiting a bridge with a neural scenario, we mirror killer lymphocytes $T_K$ and $B$ cells
to neurons and helper lymphocytes $T_{H_1},T_{H_2}$ to synapses, hence showing
that the immune system may experience the Pavlov conditional reflex phenomenon: if the exposition
to a stimulus (EBV antigens) lasts for too long, strong internal correlations among $B,T_K,T_H$ may
develop ultimately resulting in a persistent activation
even though the stimulus itself is removed. These outcomes are corroborated by several experimental findings.
\end{abstract}

\section{Introduction}

Chronic Fatigue Syndrome (CFS) refers to a
clinical condition characterized by a persistent debilitating
fatigue, neurological problems and a combination of flu-like symptoms (e.g. headache, tender lymph nodes),
ranging from at least $6$ months up to several years \cite{lancet,review1,review2,review3,review4,review5}.
\newline
The estimated worldwide prevalence of CFS is $0.4\% - 1\%$ (meaning  over $800 000$ people in the United States and
approximately $240 000$ in the UK) with a striking
socio-economical impact: The average annual total value of lost
productivity in the United States is $\$9.1$ billion \cite{eco}. Such
numbers propel a continued research to determine the cause and
potential therapies for CFS, whose diagnosis is still symptom-based and whose origin remains elusive.

The resemblance of the CFS to a chronic form of acute infectious
mononucleosis (AIM) has provoked investigation on whether this illness
(whose etiologic agent is the Epstein-Barr virus, EBV), can prompt a chronic immune reaction in the body.

In this work we try to deepen this point: By bridging between a
neural system and an adaptive immune system, we show that
an associative learning phenomenon might underlie a transition from an AIM to a CFS state.
Our model focuses on the mutual interaction and regulation
between B and T lymphocytes: when stimulated by a viral load (e.g. EBV antigens),  they
get activated (AIM phase); the active phase is then supposed to relax to a quiescent state once the viral load ceases.
Actually, we evidence that, during such a relaxing stage, the collective behavior of the components of the system
can yield non-trivial phenomena: if AIM phase takes a relatively
long time to recover (with respect to the timescale that sets the
standard immune response, say two weeks), which is in turn
related to the success of the EBV to elude
immuno-surveillance, a strong correlation between the activation of
B and T lymphocytes can be accomplished. As a result, when the
viral load has vanished and B cells remain activate for their
memory role, T cells can also maintain high concentration levels since they have ``learnt'' that active
B cells are associated to
infection\footnote{In this context there is no difference between
activation Monte Carlothrough division among plasma and memory cells \cite{abbas}
or Monte Carlothrough Couthino idiotipic/anti-idiotypic internal images \cite{tony,cazenave};
both signals are meant to work analogously.}. Therefore, as T cells display strong inflammatory properties,
we may get a state of chronically active
immune response (with CFS symptoms)  despite the original infection is no longer in
course.

We stress that our approach, by applying basic concepts of statistical mechanics to immunology,
points out \emph{emerging} possible mechanisms leading to the development of the CFS. Accordingly, we overlook the details of the interactions at work in order to focus on the very key mechanisms underlying the phenomena\footnote{In a very simplified parallel, phase transition classification in statistical mechanics evidences the
existence of abrupt macroscopic changes occurring in the system under investigation,
when varying its control parameters: although the mechanisms underlying e.g. the "ice-water"
transition and the precipitation in an acid-base titration are completely different,
the global phenomenology - described in the proper specific set of observable -
behaves in the same way and a lot of mathematics and physics can be shared in their modeling (first of all the
minimum energy and maximum entropy principles).} :
Interestingly, as we will show, for
this learning process to be properly fulfilled, T cells need to
bypass the helper signal from specialized lymphocytes and this
has been recently evidenced experimentally.

The paper is structured as follows: In Sec.$2$ we provide a basic
background about EBV, CFS and associative learning; these topics will be merged in
Sec. $3$, where we present our model. Then, in Sec. $4$, we show our
analysis and results. Finally, our conclusions and discussions are
in Sec. $5$, while all the mathematics involved is reported in the
appendix.

\section{Minimal background}

In this section we provide a basic background  about the
main features concerning chronic fatigue syndrome, Epstein-Barr virus and classical conditioning, then, in the following
sections we will merge such concepts to get an interpretation for
the emergence and establishment of the CFS.

Before proceeding, it is worth introducing the main agents of the adaptive immune system \cite{abbas},
highlighting the details crucial for our framework.

\subsection{The (adaptive) immune system}\label{sec:immuno}

An immune response is generally triggered by the introduction into the body of an \emph{antigen}, which may
have either exogenous or endogenous origin (e.g. toxins, bacteria, viruses, cancerogenic cells). For instance,
the genes of viruses that have infected a host cell can encode several proteins working as antigens.

\emph{B lymphocytes} are the agents of the humoral (i.e. mediated by secreted antibodies) immune response. B-cells
can produce antibodies upon their full activation, and are divided in clones (ensembles of cells all producing the
same antibody). Activation requires antigen recognition as well as a signal from (antigen stimulated) $T_H$ cells.
From activated B cells, specific for a given antigen, memory cells are eventually formed; these are long-life cells
able to respond quickly to a following exposure to the same antigen.

Furthermore, Jerne, in the 70's, suggested that antibodies must not only detect antigens, but also function as individual internal images of certain
antigens and are themselves detected and acted upon. Via this mechanism, an effective network of interacting antibodies is formed, where network interactions provide a
"dynamical memory" of the immune system, by keeping the concentrations of antibodies at appropriate levels. This can be understood as follows:
At a given time a virus is introduced in the body and starts replication. As a result, at
high enough concentration, it is found by the proper B-lymphocyte counterpart. Let us consider, for
simplicity, a virus as a string of information (i.e. $V = 1001001$); the complementary\footnote{The dichotomy of a binary alphabet
in strings mirrors the one of the electromagnetic field governing chemical bonds} B-cell producing
the antibody Ig1, which can be thought of as the string
$V^*= 0110110$, will then start a
clonal expansion and will release high levels of Ig1. Consequently, after a while, another B-cell
will meet Ig1 and, as this antibody never (macroscopically) existed before, it will attack it by releasing
the proper immunoglobulin Ig2 (i.e., the string $V^{**}=1001001$). The latter is actually a "copy" (internal image) of the original virus
but with no DNA or RNA charge inside.

\emph{T lymphocytes} are the agents of
cellular-mediated (i.e. not involving antibodies but directly cellular mechanisms such as lysis) response.
As in standard literature, we focus on T-helper cells ($T_H$) and on T-cytotoxic or killer cells ($T_K$) which,
when in their quiescent state, are referred to as T-CD8+ and T-CD4+, respectively. Quiescent T-cells can be activated
upon contact
with cells which have previously interacted with the antigen: T-killer cells can interact with the so-called
Class I Major Histocompatibility Complex (MHC-I) expressed by all cells, while
T-helper cells  can interact with the so-called Class II Major
Histocompatibility Complex (MHC-II) expressed only by antigen presenting cells (APC, e.g.
macrophages, dendritic cells, B-cells).
Active $T_K$ expresses killer functions destroying infected
cells, while active $T_H$ assists other white blood cells in immunologic processes,
including maturation of B cells and activation of cytotoxic T
cells.

Every immune system cell is equipped to synthesize
and release a variety of small molecules, called \emph{cytokines},
that travel to other cells (both immune and
not-immune) and up/down-regulate their growth; cytokines include
interferons (IFNs) and interleukins (ILs).

Before turning to our framework, built of by means of statistical-mechanics techniques, we stress that several
other aspects and techniques stemmed from the fields of mathematics and theoretical physics are becoming available
to investigate the biological world, ranging from kinetic theories \cite{post2}, to associative neural networks \cite{JTB}, to cellular automata and
more \cite{post1,post3}.

\subsection{Chronic Fatigue Syndrome}\label{sec:CFS}

The literature on CFS is very broad with hundreds of
analysis carried out and a rich collection of
data, yet the clinical implications of such findings remain
uncertain and a unifying, globally accepted, picture of its
etiology and pathophysiology is still missing \cite{review1,review2,review3,review4,review5}.

Current theories are looking at the possibilities of neuroendocrine dysfunction, virus geneses, environmental toxins,
genetic predisposition, or a combination of these: Several researches  suggest that Epstein-Barr Virus (EBV),
by prompting a chronic immune reaction in the body, might cause CFS.
Indeed, the phenomenology reported is consistent  with
the idea that the syndrome may follow the occurrence of an
infection yielding a massive immune response, which, for causes
not yet completely clarified, may persist for long time,
although the underlying infection is no longer in course.
In fact, a CFS state is usually associated to an abnormal concentration
and/or functioning of B-cells, T-cells and cytokines.
Another interesting and robust immunological fact found in patients with  CFS
is an unusually high (more than $67 \%$)
increase of activated CD8+ cytotoxic T lymphocytes with MHC-II activation markers \cite{phenom1,phenom2,phenom3,phenom4,phenom5}.
This will be a key point of our speculation.

From a symptomatology viewpoint, fatigue is a common symptom, but CFS is a multi-systemic disease including even
post-exertional malaise, unrefreshing sleep, widespread muscle, joint
 pain, cognitive difficulties, chronic (often severe) mental and physical exhaustion,
 muscle weakness, hypersensitivity, orthostatic intolerance, digestive disturbances and more.

\subsection{The Epstein-Bar virus}\label{sec:EBV}
EBV is one of the most successful viruses, infecting over $90\%$
of humans and persisting for the lifetime of the person in a non
pathogenic way\footnote{strictly speaking, EBV is also associated to serious diseases as Burkitt lymphoma, but its incidence is irrelevant with respect to its average behavior, which is the typical outcome from a many-body theory as statistical mechanics} \cite{ebv,ebv2}. The infection can follow different pathways, in
particular, it can turn in AIM (in up to $25\%$ cases \cite{dedra})
or it can simply introduce the virus in the host organism in
a non apparent way.

The virus aims to enter B-cells and, if successful, two outcomes are possible: In the first case the EBV
begins a viral replication cycle (so called "lytic phase'',  a common feature of most viral infections), which
induces the death of the infected cell, followed by the complete
release of new virus particles, which are going to infect other
cells; in the second case a state of latency (latent
phase) is established where the ``disguised'' virus multiplies and stands by
inside the cell, while no extracellular phenomena are observed, in such a way that no tackling by the immune system is evidenced.

During the primary infection, the latent cycle and the lytic cycle
proceed in parallel and the immune system
addresses most of its resources to the lytic cycle of viral
replication; the infection can be asymptomatic, have
non-specific symptoms, or be so massive to result
in AIM.
The acute phase can last up to several months and it ceases when the lytic cycle is interrupted by the immune responses or
by the virus itself, then, the infection becomes latent and the host
becomes a Healthy Carrier.

The possible persistence of the acute phase, despite a potent immune
response against it, indicates that the virus has evolved
strategies to elude the immune system. Among the different
hypothesis, one has received
particular attention \cite{BCRF1}: the antigen BCRF1\footnote{The BCRF1 antigen is a Lytic Antigen sharing $70 \%$ of the human IL-10R,
which is the membrane bound receptor for IL-10, see also \cite{ebv2}.} can simulate the
signal produced by IL-10 cytokines (which normally prompts leukocytes specialized against small-sized threatening
agents, like EBV's antigens) and determine a delay in the immune
response. More precisely, the signal from BCRF1 inhibits the
production of real IL-10; the lack of IL-10 polarizes the cellular
immune response in the activation of a different kind
of leukocyte, specialized in fighting against bigger-sized
pathogens.

We finally report an interesting study \cite{107} on T-cell responses, in the cases of a relatively brief (2-3 weeks) and of a protracted (4 months) acute phase. Although expansions of
antigen-specific T-cells were observed in both situations, the T-cells reactivity occurred to be broad (i.e. addressed to several, both lytic and latent, antigens) and narrowly focused (i.e. mainly addressed to a singular antigen,
the Early BMLF1), respectively\footnote{An investigation on the link between BCRF1 and BMLF1 can be found in \cite{BMLF1}}.

\smallskip

Summarizing, a significant presence of antigen BCRF1 can determine a delay in the immune response. As a result, the immune activity may take a long time for the clearance of the infection; during this time the concentration of $T_K$ cells remains high and polarizes against BMLF1 antigen as if an internal self-reinforcement has occurred.

\subsection{Statistical Mechanics of Pavlov effect}\label{sec:PAV}

Classical conditioning, experimentally demonstrated by Pavlov
\cite{pavlov}, is probably the most famous form of
associative learning. The typical procedure for inducing classical
conditioning on a subject (e.g. a dog) involves presentation of a
neutral stimulus (e.g. bell ring) along with a stimulus of some
significance (e.g. food). The neutral stimulus can be any event
that does not result in an overt behavioral response from the
subject. If
the neutral and the significant stimuli are repeatedly paired, the subject eventually associates the two stimuli and starts to produce a behavioral
response (e.g. salivation) even to the neutral stimulus alone.

From a statistical-mechanics point of view, classical conditioning can result from the interplay of dynamic phenomena, as early investigated in \cite{guerra}. More precisely, statistical mechanics usually assumes that the states of interacting ``components'' are fast variables, while coupling among them evolves on much larger time scales, in such a way that, according to adiabatic
hypothesis, the whole process results in two distinct time sectors; for instance, in neural scenario,
learning and retrieval correspond to the fast (neural) and slow (synaptic)
dynamics respectively \cite{ton}. Conversely, Pavlov phenomenon emerges when these two
timescales are not so spread and a unique, coupled temporal
evolution can be considered for retrieval variables and learning ones.

To fix ideas, let us
introduce a basic model which shall be exploited in the following.
We consider two  (on/off)-neurons $\sigma_i = \pm 1$ $(i = 1, 2)$
connected by one synapse $J = \pm 1$, so that $\{ \sigma_i, J \}
\in \{ -1, 1\}$. The characteristic time for the relaxation of the
two neurons is the same and denoted by $\tau$, while the characteristic time for the
relaxation of the synapse is $\Theta$, with $\tau \ll \Theta$. The time-averaged mean
values of these three components are $m_i(t)$ and $w(t)$, for
which $\{ m_i(t), w(t)\} \in [-1, 1]$\footnote{It is worth noting that here, coherently with the statistical
mechanics philosophy, we use the ergodic hypothesis, such that the time average of a generic lymphocyte $\sigma_i$, namely $m_i$,  can be evaluated  through the ensemble average, that is averaging over the population of the identical lymphocytes at a given time $t$. Clearly, within the former approach, such averages must be taken over a time range at least order of $\tau$ to be meaningful, allowing in this way the fast relaxation mode to operate.}.
This system shows the capacity of a dynamical learning in the
following sense: consider the action of two external signals,
$(s_1, s_2)$, each applied on a different neuron
$(\sigma_1,\sigma_2)$. If the stimulation of both neurons
happens for a short time $t$, namely comparable with the short timescale (i.e., $t \sim \tau$), once one signal is removed, the
corresponding neuron stops its activity; conversely, if the two
stimuli are presented for a sufficiently long time (i.e. $t \sim \Theta$), due to synaptic contribution,
correlations within the system develop and, if one signal is turned off, its
corresponding neuron remains active: We will refer to this
dynamical feature as associative learning.

Let us deepen in more technical details the emergence of such a phenomenon.
At first, both signals $s_1=1,s_2=1$ are applied to the related neurons [regime $(1,1)$], consequently, the synapse
can be enforced (according to Hebb's prescription
\cite{hebb}), that is $w(t)$ grows in time. After a given time $t_s
\in [0, \Theta]$ one signal, say $s_2$, is removed [regime $(1,0)$]: if
an associative
learning is accomplished, we expect that under the action of $s_1$ alone,
the system is still able to stimulate even $\sigma_2$.
As we are going to show, these words can be translated into a
system of stochastic differential equations describing the evolution
of the neural configuration.

Let us now outline our mapping.
Neurons are $B$ and $T_K$ lymphocytes responding to antigenic
load by EBV and the synapses joining them is the helper lymphocyte tuning their activity.
All the lymphocytes producing the same antibody (BCR) for the Bs, or the same TCR for the Ts,
are grouped into clones, whose size can be in principle (e.g. under an insult) extremely
large. It should be underlined that, while neurons and synapses are "single" cells,
the interaction in the immune level acts at the level of clones, hence,
while single lymphocytes within each clones are
clearly single cells (showing discrete symmetries of activity $\pm 1$, as neuronal models),
their clonal behavior can be described by a continuous variable.
\newline
Now, due to the ergodic hypothesis underlying our statistical-mechanics approach,
we can interchange the average over time with the average over the ensamble of all the $N$ lymphocytes $\sigma_i^{(j)}, j=1,...N$, belonging to the clone considered.
 Namely, instead of evaluating the time integral of $\sigma_i(t)$ to obtain its typical behavior $m_i$, we average over the ensemble of all the same lymphocytes (hence the B-clone under expansion for the B, or the killer or helper for the Ts). This has the advantage of turning the evolution described in terms of Markov chains into Langevin stochastic dynamics, whose integration can be accomplished easily through standard numerical techniques.
In the following, we display only the evolution of the averages $m_1=\langle \sigma_1 \rangle$, $m_2=\langle \sigma_2 \rangle$ and $w=\langle J \rangle$, described by the system of differential equations (see Appendix for more details on its derivation):
\begin{eqnarray} \label{eq:system}
\tau \frac{d m_1}{dt} &=& - m_1 + \tanh\left[ \beta (w m_2 + s_1) \right], \\
\Theta \frac{d w}{dt}  &=& - w + \tanh\left( \beta m_{1}m_2 \right), \\
\tau \frac{d m_2}{dt} &=& - m_2 \tanh\left[ \beta (w m_1 + s_2) \right].
\end{eqnarray}
The randomness in the stochastic evolution is ruled by $\beta \in \mathbb{R}_0^+$, which encodes the level of noise in
the system such that for $\beta=0$ the dynamics is completely
random (coherently  the observables average to zero as they are
symmetrically distributed), while for $\beta \to \infty$ the
hyperbolic tangent becomes the sign function and the dynamics is
completely deterministic.

Finally, we stress that the statistical mechanics model we have elaborated allows a formal picture of phenomena which, actually, go far beyond Pavlov's conditional reflex; more generally, it describes processes of associative learning which has been evidenced in different biological contexts \cite{assobio}.

\section{Reading the CFS puzzle from a Pavlov perspective}
Recalling the evidence of  two
time-scales characterizing the evolution of EBV primary infection (see \cite{107} and
Sec.~\ref{sec:EBV}), as well as the statistical mechanics model
introduced in Sec.~\ref{sec:PAV}, we want to exploit the concept
of associative learning as a bridge between the neuronal  and the
immune contexts; the occurrence of such a learning process might
be interpreted as a cause for the establishment of a CFS (see
Sec.~\ref{sec:CFS}).

The core of our mapping is that the typical timescale for an effector response (B/K) can be smaller than the AIM in the long lasting case, hence allowing the development of correlation between the two branches. The last point could allow them to bypass helper authorization to clonal expansion. Of course, the statistical mechanics approach does not provide an explanation at molecular level (being unaffected by molecular details), but it suggests that in this context such outflanked helpers enable the genesis of such correlation acting as a positive (reinforcing) point of contact among them.

Let us summarize the evolution of the phenomenology of the EBV
pathology within this perspective in three main phases, where we also anticipate the mapping between the antigenic load on $T_K$ and on $B$ cells respectively and the state of
the two stimuli ($s_1,s_2$):

\begin{figure}[tb] \begin{center}
\includegraphics[width=.5\textwidth]{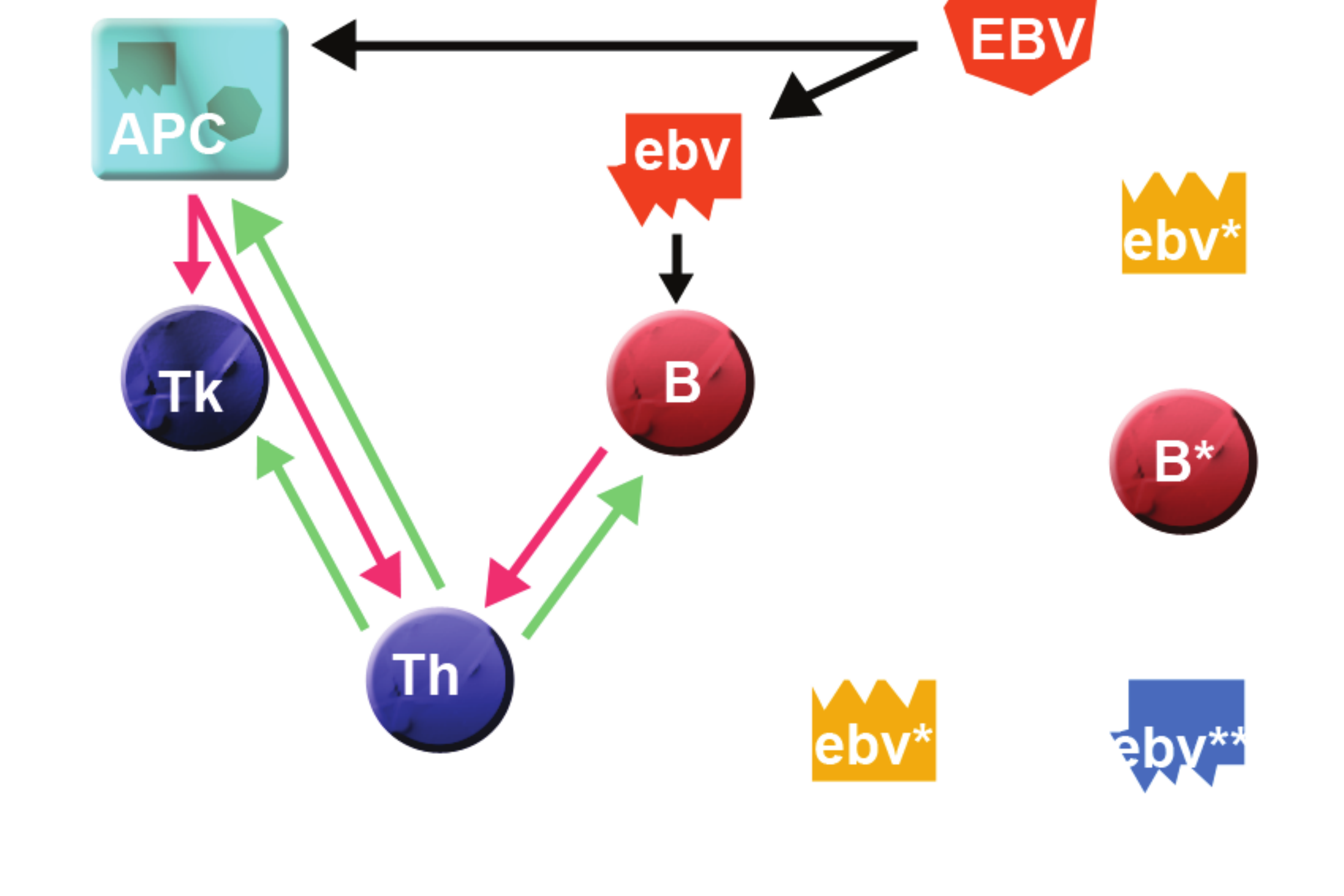}
\caption{\label{fig:uno} Circlets represent lymphocytes
(T-cells in blue, B-cells in red); the red pentagon represents the Epstein-Barr Virus; irregular shapes are antigens (red), antibodies ebv*
(yellow), or anti-anti-body ebv** (azure); the rectangular shape represents
APCs or infected cells (healthy B-cells are not included). Different stages of the process are depicted in different colors: Infection (black arrows), Presentation (pink arrows), Activation (green arrows).}
\end{center}
\end{figure}

\begin{itemize}
\item
\emph{Infection and presentation}, [Regime $(0,0)$]. The
virus starts the Lytic Cycle and
infects permissive cells, producing antigens. The infected cells (via MHC-I) and the APCs (via MHC-II) present the processed antigen for specific T lymphocyte's
recognition. A resting CD4+ T-cell must be triggered via MHC-II present on either an APC or a (antigen-specific)
 B-cell (pink arrows in Figure \ref{fig:uno}) to become
an activated $T_H$.
\item
\emph{Activation and response}, [Regime $(1,1)$]. As the virus load overcomes the low dose tolerance of the lymphocytes, B-cells become fully activated by $T_H$
(green arrow in Fig.~\ref{fig:uno}), so that they can now start to differentiate into memory
cells and plasma cells producing antibodies; $T_H$ cells also provide the second signal, indirectly via APC, for a resting
CD8+ T-cell. A resting CD8+ T-cell must be triggered by an APC via MHC-I
(first signal, pink arrow in Fig.~\ref{fig:uno}) and by $T_H$
(second signal, green arrow in Fig.~\ref{fig:uno}) in order to become a $T_K$.\\
The activated $T_K$ can now operate directly (yellow arrow in Fig.~\ref{fig:due}) on APC
that is still in contact with the $T_H$, or that has been
previously instructed. Meanwhile, the B-cell starts to produce
antibodies (ebv*); such antibodies, being actually new antigens for the
immune system, undergo an equivalent recognition
process (yellow pattern in Fig.~\ref{fig:due}), so that they stimulate
their specific B*-cells for T-cell presentation \footnote{This standard mechanism of anti-antibodies production is due to the fact that the antibodies produced constitute a large concentration of proteins seen as anomalous by the host itself. We
stress that, for this mechanism to hold, we do not need the Jerne
idiotypic cascade, but only the Coutinho internal image, which has
been largely revealed experimentally \cite{tony,cazenave}.}.
\item
\emph{Possible Learning}, [Regime $(1,0)$]. The immune response eventually annihilates the antigenic load,
the viral load ceases and $T_K$ cells, no longer stimulated, can undergo apoptosis.
As for $B$ cells, they maintain a certain degree of stimulation (memory) even though
of different nature with respect to the original one. In fact, as mentioned in Sec. \ref{sec:immuno},
after activation, B* cells secrete antibodies (ebv**) that, being ``complementary of the complementary'' \cite{cout}, resemble the
original virus (ebv); as a result, they may act as signals themselves, that is,
the antibodies ebv** sustain the
stimulation of B cells \footnote{A more simplified description
would require only the presence of B-memory cells for providing
signalling to $T_H$, skipping any discussion on memory generation
in $B$-cell network \cite{cout}.}.
\newline
\begin{figure}[tb] \begin{center}
\includegraphics[width=.5\textwidth]{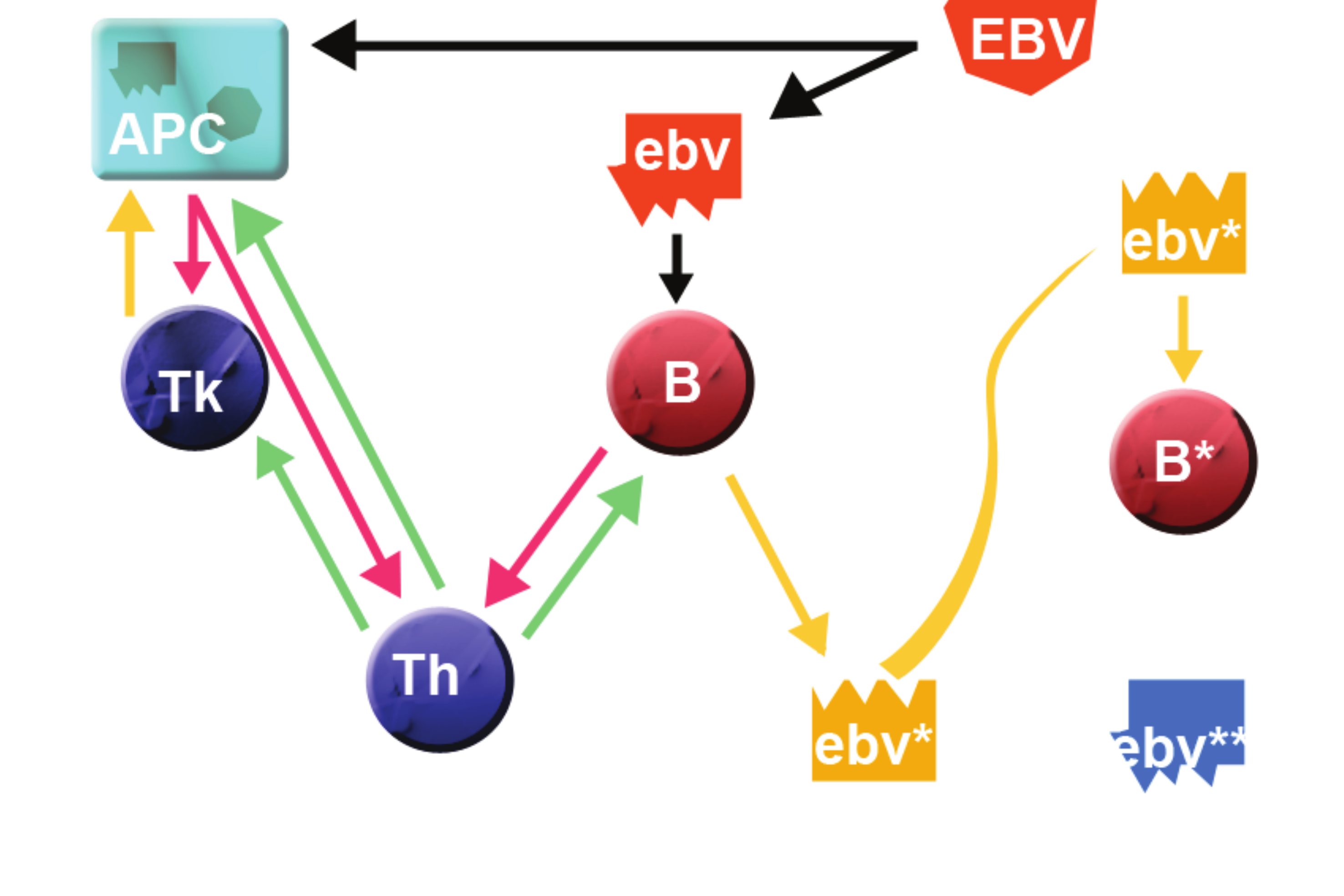}
\caption{\label{fig:due} Kill and Memorize (yellow arrows). Agents are the same as in Fig.~\ref{fig:uno}}
\end{center}
\end{figure}
Now, according to the duration of the co-stimulation [Regime $(1,1)$] two alternative situations
might happen:\\
\begin{itemize}
\item
\emph{Healthy Carrier State, HCS.} If the Lytic Cycle has been
interrupted within a relatively short time $t_s$ by the immune response,
no associative learning between the production of $T_K$ cells and the
production of B-cells is accomplished. The immune system
has stored memory of the infection via memory cells and
the EBV latency has established. The patient becomes a Healthy
Carrier displaying specific memory healthy cells  for those given
antigens, as well as infected resting B cells in the Latent Cycle.\\
\item
\emph{Chronic Fatigue Syndrome, CFS.}  The prolonged exposition to the (original) viral load
(experimentally found to occur in the presence of large concentration of BRCF1 antigen) can lead to
an associative learning
between $T_K$ and B cells production.
In fact, although $T_K$ are no longer directly stimulated by the antigen, the active state of $B$ cells
can work itself as a surrogate stimulus. Namely, $T_K$ bypass the $T_H$ signal and interact directly with
the MHC-II signal provided by B-cells as APC. This is consistent with \cite{107}, where it is reported that the
BMLF1-specific CD8+ T-cell (which should only recognize the class
MHC-I) gets active bypassing the necessary T-CD4+
indirect signal. This scenario would lead to a chronic activation
state and it will be further discussed in the next Section\footnote{As already stressed, in our
model there are no suggestions for this MHC-I/MHC-II switch, as these details of the interaction are not even introduced. On the other hand, the unbalanced $T_K$/$T_H$ load emerging from our stochastic dynamics can be explained through this mechanism.
In  this sense, statistical mechanics, evidencing key mechanisms, provides hints for understanding experimental findings.}.
\end{itemize}
\end{itemize}

\begin{figure}[tb] \begin{center}
\includegraphics[width=.45\textwidth]{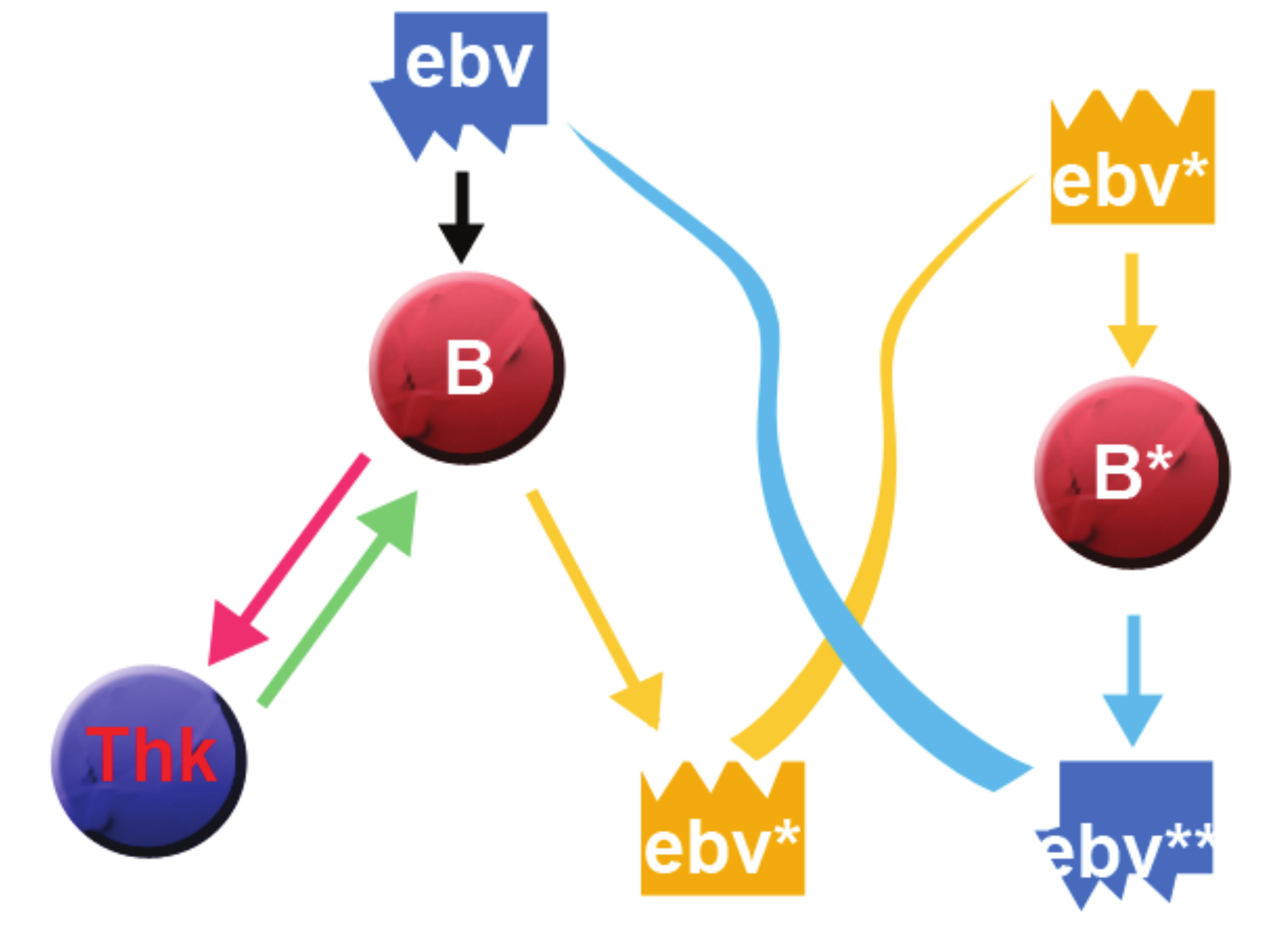}
\caption{\label{fig:tre} Learning. Agents are the same as in Fig.~\ref{fig:uno}}
\end{center}
\end{figure}

To summarize, this is our proposal for the CFS etiology:  a massive presence of BMLF1 antigens makes the
 clearance of the infection slow so that a long co-stimulation of BMLF1-specific B and $T_K$ cells takes place. During
 this stage a learning process occurs making $T_K$ cells able to detect the signal directly from $B$ cells,
 hence by-passing the direct stimulation from the antigen as well as from $T_H$. One can think at this situation as if $T_K$ assumed both killer and helper functions, that is as if it switched to a hybrid state.

In order to corroborate our speculation, we studied the dynamical properties of our model through analytical arguments and numerical simulations.

\subsection{Formalization}
In our interpretation, the viral load represents the external signal; when $s_i =
+\infty$\footnote{The choice of this limit value is for
simplifying calculations, the physics behind is essentially the
same of every "high enough" load. We stress, however, that the value
of the field, as introduced into our stochastic system, is not
coupled to the noise, such that a high value implicitly accounts
for low noise (i.e. high $\beta$).}, the viral load is much bigger
than threshold levels implied by low-dose tolerance \cite{BA2,BA3},
conversely, when $s_i = 0$, it is much lower. $T_K$ and $B$ cells belonging to clones specific for EBV antigen play as the
neurons $\sigma_1$ and $\sigma_2$; the synapse $J$ represents the $T_H$ cell.
Indeed, the $T_H$ can influence both $T_K$ and $B$ lymphocytes,
via the sub-populations $T_{H_1}$ and $T_{H_2}$, respectively.
Hence, the "synapsis" should be thought of as a proper combination
of $T_{H_1}$ and $T_{H_2}$, which results in a long relaxation
time $\Theta$ (see Sec.~\ref{sec:parallelo}).

Following \cite{BA2,BA3}, the real  size of the clone can be related
to the time-average value of the representative spin, bounded in $[-1,1]$, by means
of an exponential law:
\begin{equation} \label{eq:real_size}
M_i =  \exp \left[ \eta_i \left(\frac{m_i(t)+1}{2} \right) \right],
\end{equation}
where $i=1,2$, and
\begin{equation}
W = \exp \left[ \eta_0 \left(\frac{w(t)+1}{2} \right) \right],
\end{equation}
being $\eta_i$, $i=0,1,2$ a parameter that introduces the size of the pertaining population.
These positions reproduce the chemical kinetics equation for the concentrations.
Here we reasonably assume that the clones considered can be bounded by the same size $N = 2 \times
10^{11}$, so that $\eta_i = \log(N)$ \cite{abbas}.

Now, the AIM phase corresponds to a Regime $(1,1)$,  where both
$T_K$ and $B$ clones are stimulated by EBV antigens.
The AIM phase is estimated to last from two weeks up to two months,
so we choose $\tau$ in the interval $\tau \in  [14; 60]$ days.

Once the viral shedding has been interrupted by the immune
responses, the $T_K$ clone
is no longer stimulated, while the B-cell clone is still managing
the memory of the infection (directly or Monte Carlothrough its conjugate
specific antigen): This corresponds to the Regime $(0,1)$.
\newline
The two-steps evolution described here is a useful  schematization
for the analytical approach developed in the Appendix for the
special cases $\beta=0$ and $\beta \rightarrow \infty$. More
generally, the system of coupled differential equations (\ref{eq:system})
can be solved numerically for any value of $\beta$ and in the
presence of continuous signals. In particular, while $s_2$ is still non-null over the whole range considered, $s_1$ can be chosen as exponentially decaying (being $t_s$ characteristic time for vanishing), in agreement with experimental findings \cite{abbas}.
\newline
Despite the system is well described by the stochastic dynamical
equations (Eqs.~(\ref{eq:system})), we can improve the picture by including proper terms
accounting for the collective behavior due to interaction with other lymphocytes and immune agents,
that is, we mimic both quiescence induction and internal signaling to apoptosis
by introducing
further two small negative fields $|\epsilon_1|, |\epsilon_2| \ll
1$, in such a way that the effective fields are
$\tilde{s}_1 = s_1 + \epsilon_1$ and $\tilde{s}_2 = s_2 +
\epsilon_2$, respectively. We underline that the statistical mechanics reason for these small fields, whose effect would otherwise be negligible being them infinitesimal,
is only breaking the gauge symmetry of the model, so to allow a quiescent state in the absence of signals.

\subsection{On the mixed synapse and timescales} \label{sec:parallelo}

In our model the helper T-cell plays the role of the synapse. Upon
activation, helper T cells differentiate in two major subtypes
 known as $T_{H_1}$ and $T_{H_2}$: Beyond other functions, the former
maximizes the proliferation of cytotoxic CD8+, the latter
stimulates B-cells into proliferation; also, they both produce
cytokines which are aimed to their own proliferation and
cross-regulate each other's development and activity \cite{abbas}.
The net result is that, once the $T_H$ response begins to develop,
it may get polarized in one of the two directions (either Type 1 or
Type 2), due to auto-amplification and cross regulation \cite{abbas}.

Now, when the two sub-populations are completely balanced and small (corresponding to a quiescent state) none of the two prevails
so that, in our equivalent model, there is no link between $B$ and
$T_K$ ($w=0$) and no learning can be established among them (as intuitively $\Theta \to \infty$).
\newline
In general, one can assume that the characteristic timescales of
lymphocyte growth are the same, independently of the particular
type, that is, $T_H$, $T_K$ and $B$ cells require the same time $\tau$ to adjust their concentrations responding to a signal.
However, in our model, the three agents considered feature
different degrees of complexity: while $B$ and $T_K$ can be thought of as homogenous populations, $T_H$ displays inner degrees of freedom, being the combination of the
two sub-populations $T_{H_1}$ and $T_{H_2}$. As a consequence, as the dynamics of $J$ tunes the unbalance between these populations (and they grow closely), its relaxation time $\Theta$ is assumed larger than $\tau$.

\begin{figure}[tb] \begin{center}
\includegraphics[width=.5\textwidth]{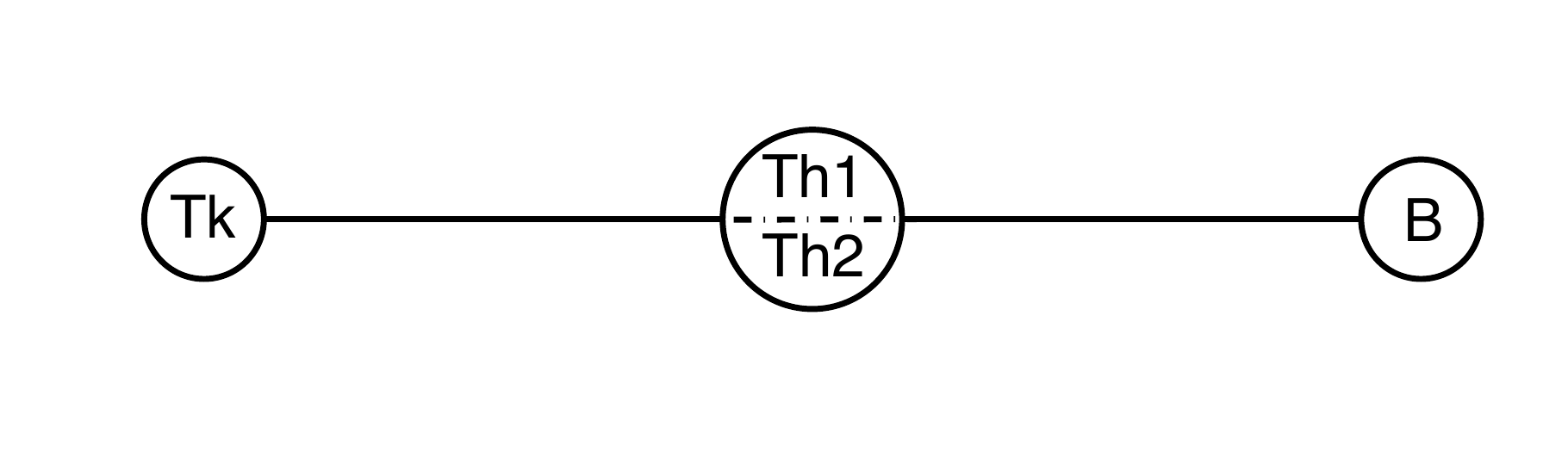}
\caption{\label{fig:parallelo} Schematic representation of the three-spins system considered, where the complex structure of the synapse (central spin) is explicitly envisaged.}
\end{center}
\end{figure}

Conversely, when
one of the two prevails the synapse is onset ($w\neq 0 $), so that
$B$ and $T_K$ can (indirectly) interact (still retaining a large timescale $\Theta$ compared to the single clone one $\tau$).
\newline
Hence, as envisaged by the scheme in  Fig.~\ref{fig:parallelo} the
central spin, playing the role of the synapse, can be thought of
as a combination of two sub-populations. An effective way
to relate the states of $T_{H_1}$ and $T_{H_2}$ with the overall state of $T_H$ is given by the following ``average'', where we denote with $w_1$ and $w_2$ the ``magnetizations'' corresponding
to the two sub-populations:
\begin{equation}
w = \frac{w_1 w_2}{4} [w_1 + w_2 + |w_1| +|w_2| + |w_1 - w_2| ].
\end{equation}
Indeed, this combination is in agreement  with the immunological
phenomenology and consistent with the model we are introducing:
balanced, quiescent sub-populations correspond to $w=0$, a large
unbalance in favor of any of the two sub-population means $w<0$, while when they are both
stimulated we have a reinforcement effect ($w>0$).

The characteristic time for the response of $B$ and $T_K$ is taken
to be $\tau \in [14, 60]$, since the AIM phase is estimated to
range from two weeks up to two months, while for our synapse we
explore $\Theta$ starting from $\Theta \geq 90$ days (as the tunable
parameter is the ratio $\tau/\Theta$, the previous choice does not
modify the results).
Furthermore, while the signal $s_2$ is constant, the  signal $s_1$
(representing the real antigenic load) is taken exponentially decaying, in such a way that the effective time of its
offset is $t_s$; similarly, in the analytical approach in the
appendix the signal is active for a time $t \in [0, t_s]$, while
the second regime holds for $t \in [t_s, \infty)$, with $\tau <
t_s < \Theta$. In any case we find that the value of $t_s$ crucially determines the final
equilibrium state.

\subsection{The role of the latent and lytic cycles}

\begin{figure}[tb] \begin{center}
\includegraphics[width=.5\textwidth]{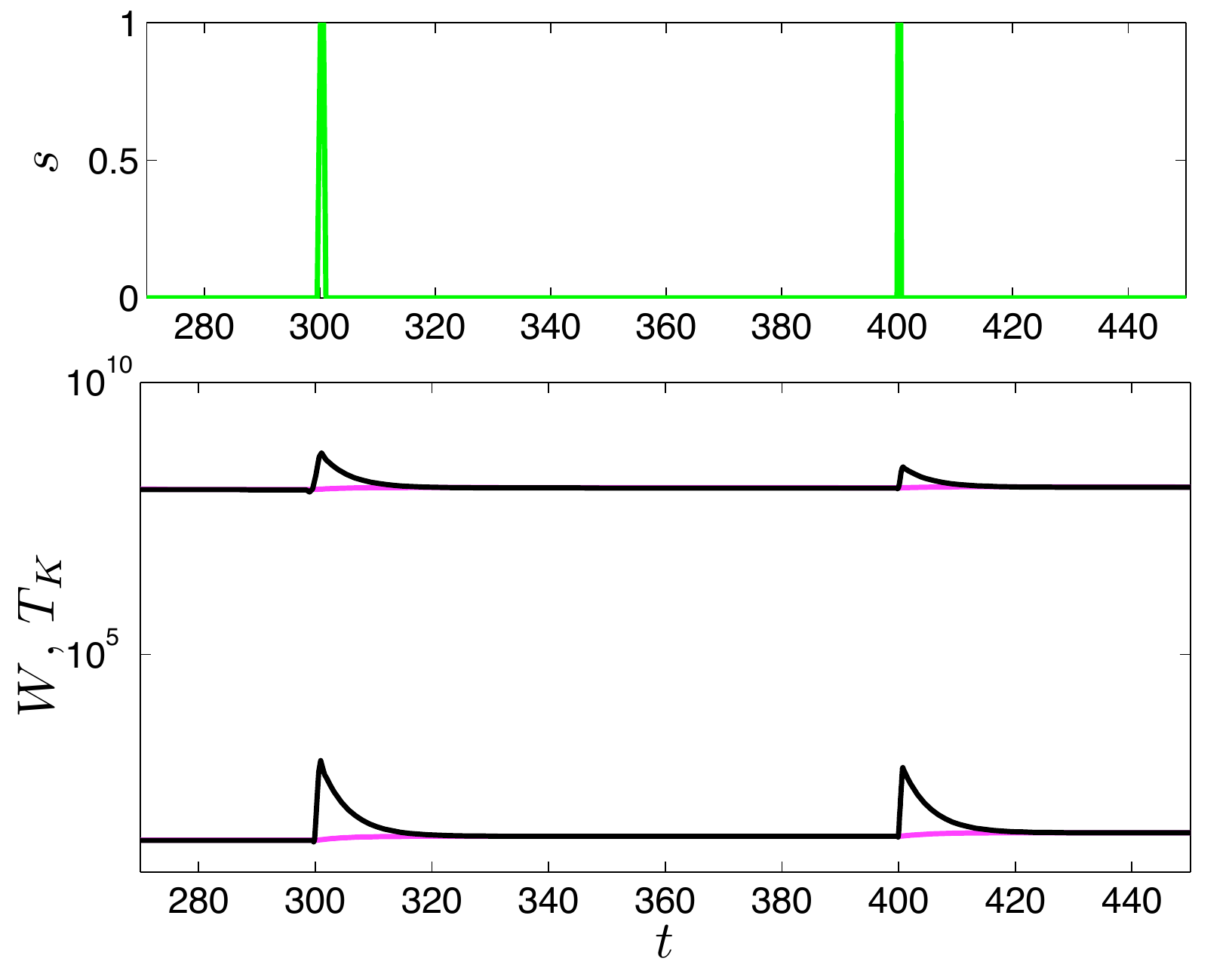}
\caption{\label{fig:impulsi} Upper panel: impulsive stimulus $s=s_1=s_2$ as a function of time; Lower panel: immune response in terms of concentration of $T_K$ (dark line) and of $T_H$ (bright line) for an healthy carrier (smaller values) and a CFS patient (higher values).}
\end{center}
\end{figure}

In this section we want to deepen a way (typical of all herpes-viruses) which may further contribute to lengthen the hospitalization time of a CFS patient.
In fact EBV, once the infection has been established in the host body, may hide away from immune
recognition and opportunely  switch between cycles of latency and cycles of lytic replication,
somehow mirroring a switch between a quiescent and an active
external stimulus (antigenic load)  acting on the immune system.
In fact, during latency, EBV mainly manages minimal  tasks as inhibiting apoptosis and blocking viral lytic replication, while, during the lytic phase, EBV syntetizes proteins from many viral genes, allowing for
nucleotide biosynthesis, RNA processing, viral DNA replication, etc.
\newline
As a consequence, within the framework based on "Pavlov phenomenology" we are using to explain the transition from an AIM to a CFS scenario, these re-activations display significantly different outcomes in healthy carriers and in CFS patients. In fact, as shown in Fig.~\ref{fig:impulsi}, for the formers, whose infection walked off quickly (toward an healthy carrier final state), sequential impulsive stimuli do not have particular consequences, while, for the latters, whose infection has been prolonged enough to allow the correlation via helpers, basically each time there is an impulsive reactivation, this thwarts the natural de-learning.

\section{Results}
The stochastic  system of Eqs.~(\ref{eq:system})-(3) is solved numerically by
means of standard Runge-Kutta packages for Matlab.
For the sake of clearness, a fully analytical solution of the system is shown in the appendix, nonetheless, here it is worth introducing the whole set of observables we need to consider. In particular, we have to deal with averages and correlations:
$$
w = \langle J \rangle_{\tau}, \ m_1 = \langle \sigma_1 \rangle_{\tau}, \ m_2 = \langle \sigma_2 \rangle_{\tau}, m_{12} = \langle \sigma_1 \sigma_2 \rangle_{\tau},
$$
$$
m_{01} = \langle J \sigma_1 \rangle_{\tau}, \ m_{02} = \langle J \sigma_2 \rangle_{\tau},
\ m_{012} = \langle J \sigma_1 \sigma_2 \rangle_{\tau}, \
$$
where the average $\langle \cdot \rangle_{\tau}$, is meant over the time and over the ensamble due to ergodic hypothesis as explained in Sec.~$2.4$ (see also the appendix).
We set our initial conditions $\mathbf{m}(0)$ using concentrations expressed
in cells/$\mu$L and keeping in mind that, in a healthy body, a
given clone has an incidence of $1$ over $10^5$ cells with respect
to the whole population; by using the normal values
for lymphocytes concentrations and the relative translation in
terms of magnetizations (see Eq.~\ref{eq:real_size}) reported in Tab.~$1$, we get:
\begin{equation}
\mathbf{m}(0)= \left\{
\begin{array}{l}
w(0) \sim 0\\
m_1(0)=-0.971\\
m_2(0)=-0.938\\
m_{12}(0) = m_1(0) \cdot m_2(0)\\
m_{01}(0) = 0\\
m_{02}(0) = 0\\
m_{012}(0) = 0
\end{array} \right.
\end{equation}
Notice that the initial value for the correlation  $m_{12}$ as the
product of the two concentrations $m_1,m_2$ is a useful condition
 to initialize the evolution (implicitly assuming un-correlation), and we will use it during
 numerical integration.

\begin{table}[ht] \label{tab:1}
\caption{Data for initial conditions \cite{abbas}. Notice that we do not specify the relative concentrations of $T_{H_1}$ and $T_{H_2}$,while we assume that their sum  $T_H$ is, at rest, zero, i.e. $w \sim 0$, meaning a balance for the two sub-populations.} \centering
\begin{tabular}{c c c c c} \hline\hline Agent & Concentration & $\%$ & $M_i(0)$ & $m_i(0)$ \\
& (cells/$\mu$L) & & (specific cells/$\mu$L) & (adimens.) \\
[0.5ex]
\hline $T_H$ & 1000.50 & 46 & 0.010 & -0.815 \\ $T_K$ & 413.25 & 19 & 0.004 & -0.971 \\  $B$ & 500.25 & 23  & 0.005  & -0.938\\ [1ex]
 \hline
 \end{tabular}
 \label{table:nonlin}
 \end{table}

In order to recover the two cases of HSC patient and CFS patient, as reported in Sec.~\ref{sec:EBV}, we consider two different
situations, corresponding to a short ($t_s=5$ days) and to a long
($t_s=100$ days) AIM phase, respectively.
The other parameters holding for both patients are:
$$
\tau=30, \ \Theta =90,
$$
$$
s_1=1000, \
s_2=e^{-(t-t_s)/80},
$$
$$
\epsilon_1=-1.0, \
\epsilon_2=-1.0.
$$
We also fix the level of noise as low ($\beta=6.5$), while later we will discuss the case of high noise.

\begin{figure}
\begin{center}
\includegraphics[width=.6\textwidth]{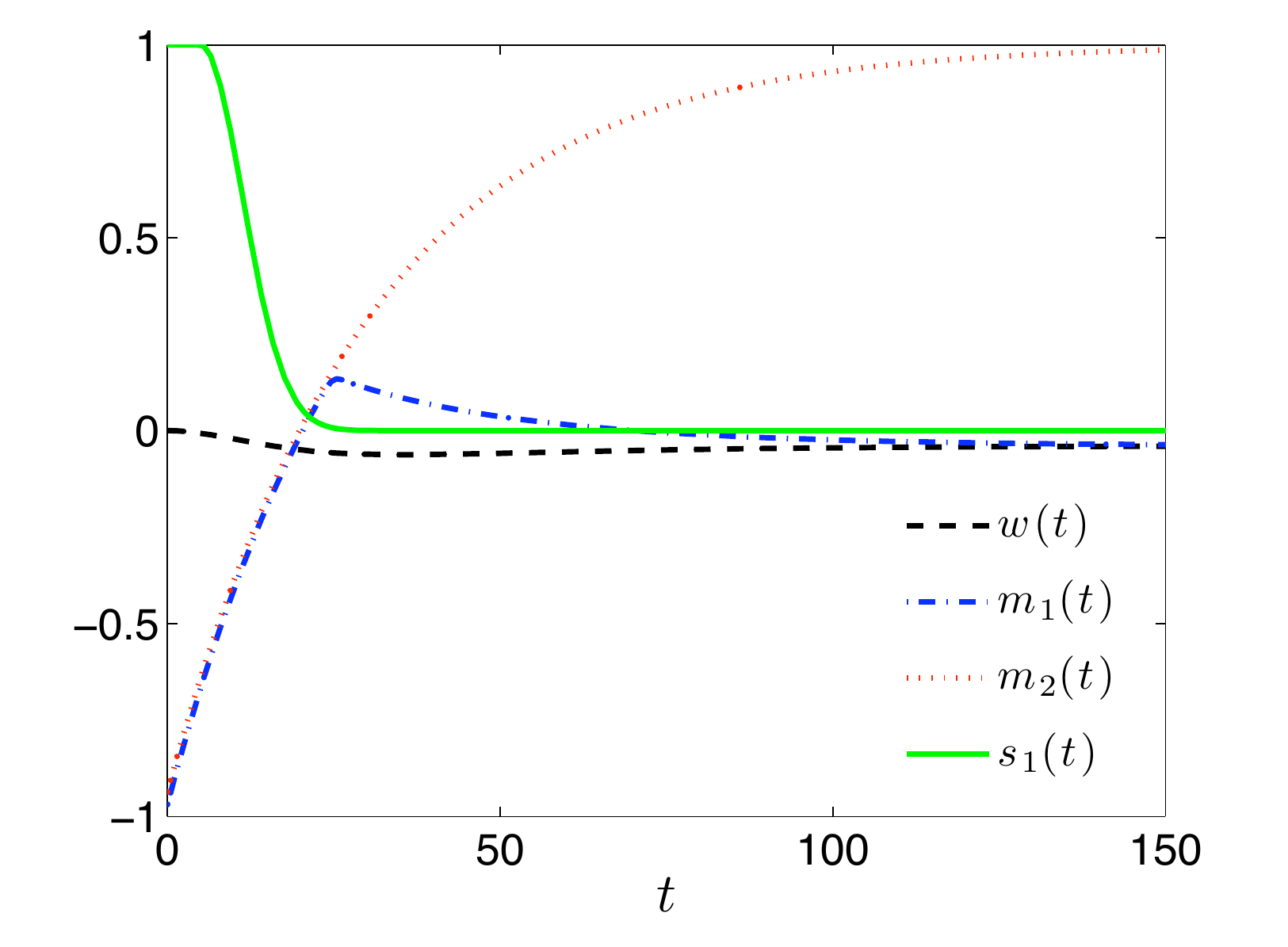}
\includegraphics[width=.6\textwidth]{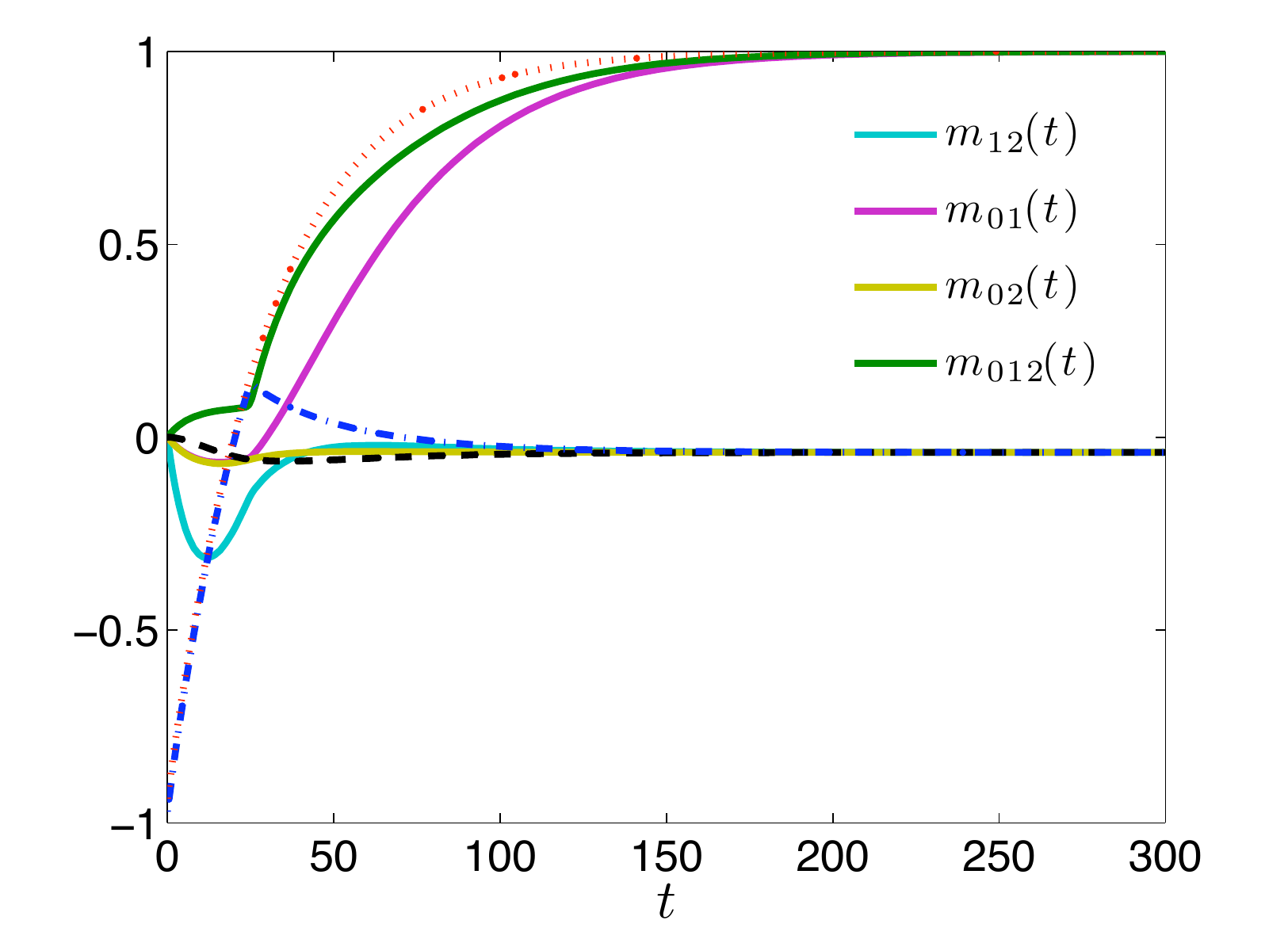}
\caption{\label{fig:short} (Color online) Top panel: Averages $w$, $m_1$ $m_2$ and signal $s_1$ as a function of time (zoom on the early regime). Bottom panel: the same averages and their correlations as a function of time, as shown in the legends; time is measured in days. The signal on $T_K$ vanishes at around time $t_s=5$. The level of noise is
low and fixed at $\beta=6.5$.}
\end{center}
\end{figure}

\emph{Patient 1: HSC Scenario.}\\
As shown in Fig.~\ref{fig:short}, during the AIM phase ($t<t_s = 5$ days),
both signals $s_1$ and $s_2$ are
active and, as  responses, $m_1$, $m_2$ grow up, meaning that $T_K$
and $B$ clones are proliferating; the synapse also increases as both its afferent inputs are growing (correlations begin).
\newline
As the real viral load is diminishing, $T_K$ concentration
decreases and finally reaches a value comparable with the initial
one. Conversely, $B$ clones, being still stimulated, maintain high levels of concentration.

\smallskip

\begin{figure}
\begin{center}
\includegraphics[width=.6\textwidth]{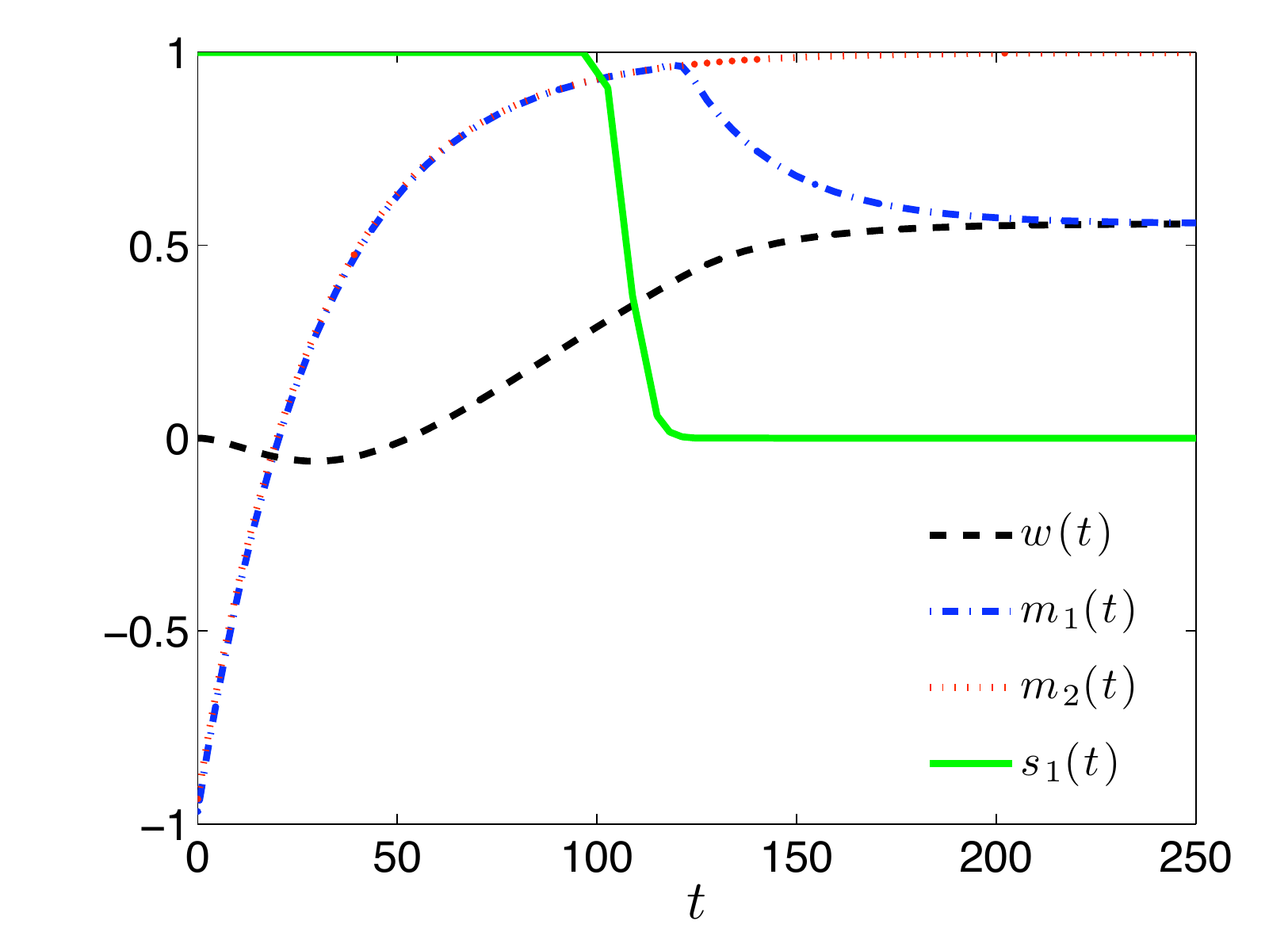}
\includegraphics[width=.6\textwidth]{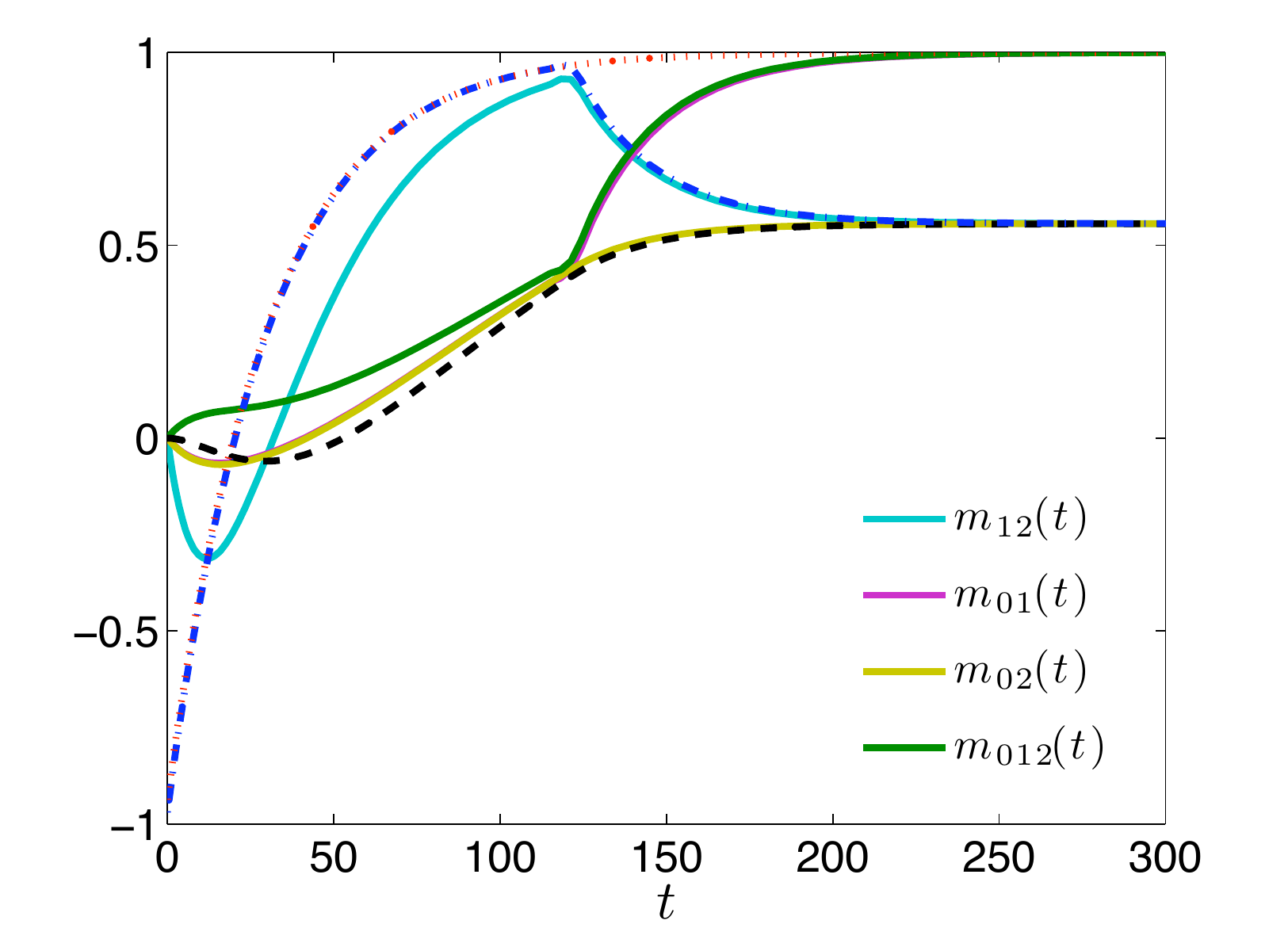}
\caption{\label{fig:long} (Color online) Top panel: Averages $w$, $m_1$ $m_2$ and signal $s_1$ as a function of time (zoom on the early regime). Bottom panel: the same averages ad their correlations as a function of time, as shown in the legends; time is measured in days. The signal on $T_K$ vanishes at around time $t_s=100$. The level of noise is
low and fixed at $\beta=6.5$.}
\end{center}
\end{figure}

\emph{Patient 2: CFS Scenario.}\\
As shown in Fig.~\ref{fig:long}, during the AIM phase ($t<t_s=100$ days), the concentrations of $B$,
$T_K$ and $T_H$ increase as a result of a viral load ($s_1,s_2
>0$). This time, the signal on $T_K$ lasts long enough for $T_K$
and $T_H$ to reach high levels (both $\sim 10^5$ cells/$\mu$L)
and, even when the signal is switched off their concentrations are
much larger than the one pertaining to Patient 1.

\begin{figure} \begin{center}
\includegraphics[width=.6\textwidth]{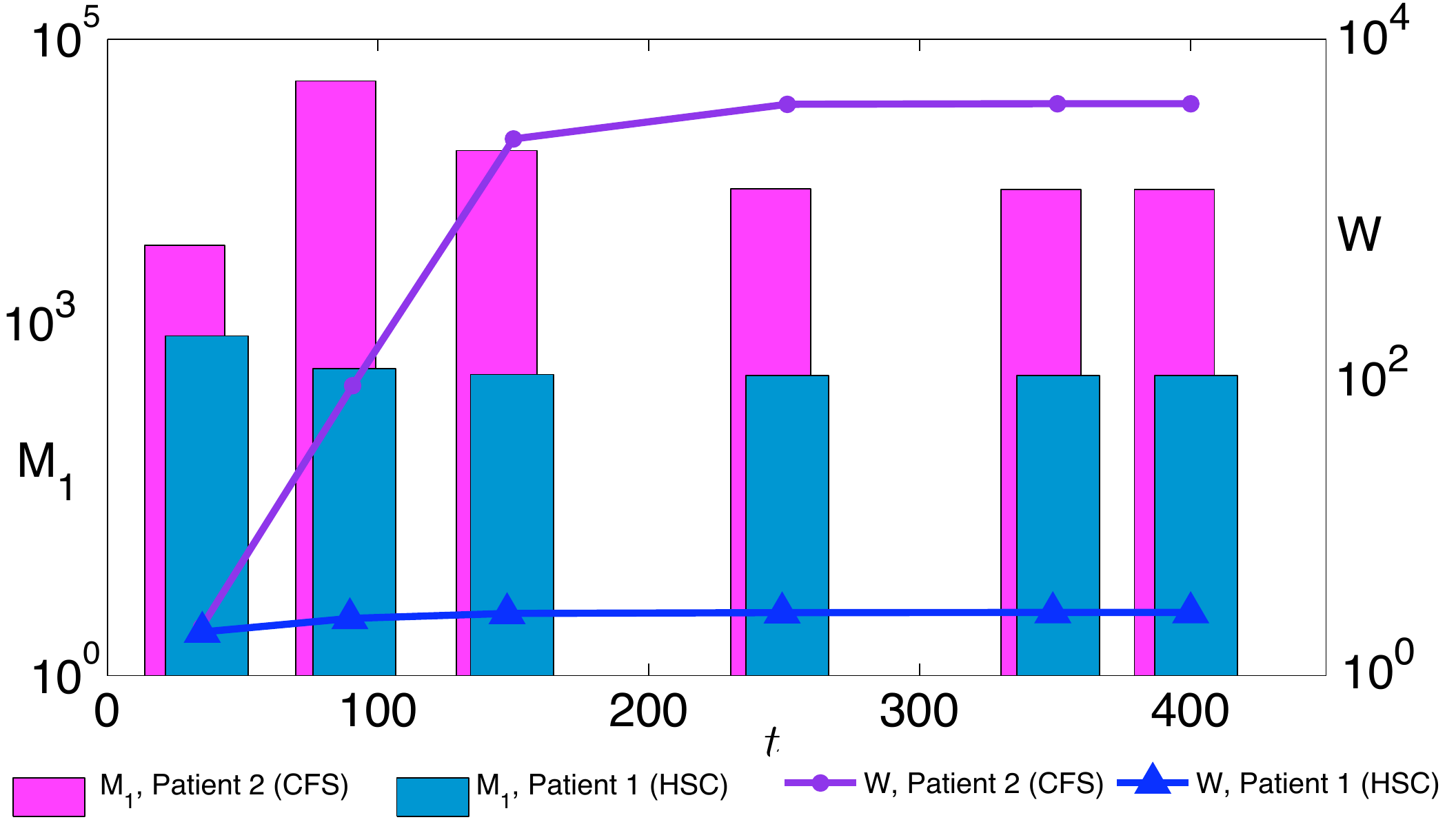}
\caption{\label{fig:compare} (Color on line) Evolution of the concentrations  of $T_K$
cells (histogram, left vertical axis) and of $T_H$ cells (curves, right vertical axis) for the two patients
considered in the case of small noise.}
\end{center}
\end{figure}

The outcomes for the two cases are compared in Fig.~\ref{fig:compare}, in terms of concentrations of  $T_H$ and $T_K$. In particular, for Patient $1$, $T_K$ displays a maximum ($\sim
10^3$ cells/$\mu$L) at very short times and then relaxes to
approximately $10^2$ cells/$\mu$L, while, for Patient $2$, $T_K$ reaches a higher maximum at longer times  and then relaxes to
approximately $10^3$ cells/$\mu$L.

\begin{figure}
\begin{center}
\includegraphics[width=.6\textwidth]{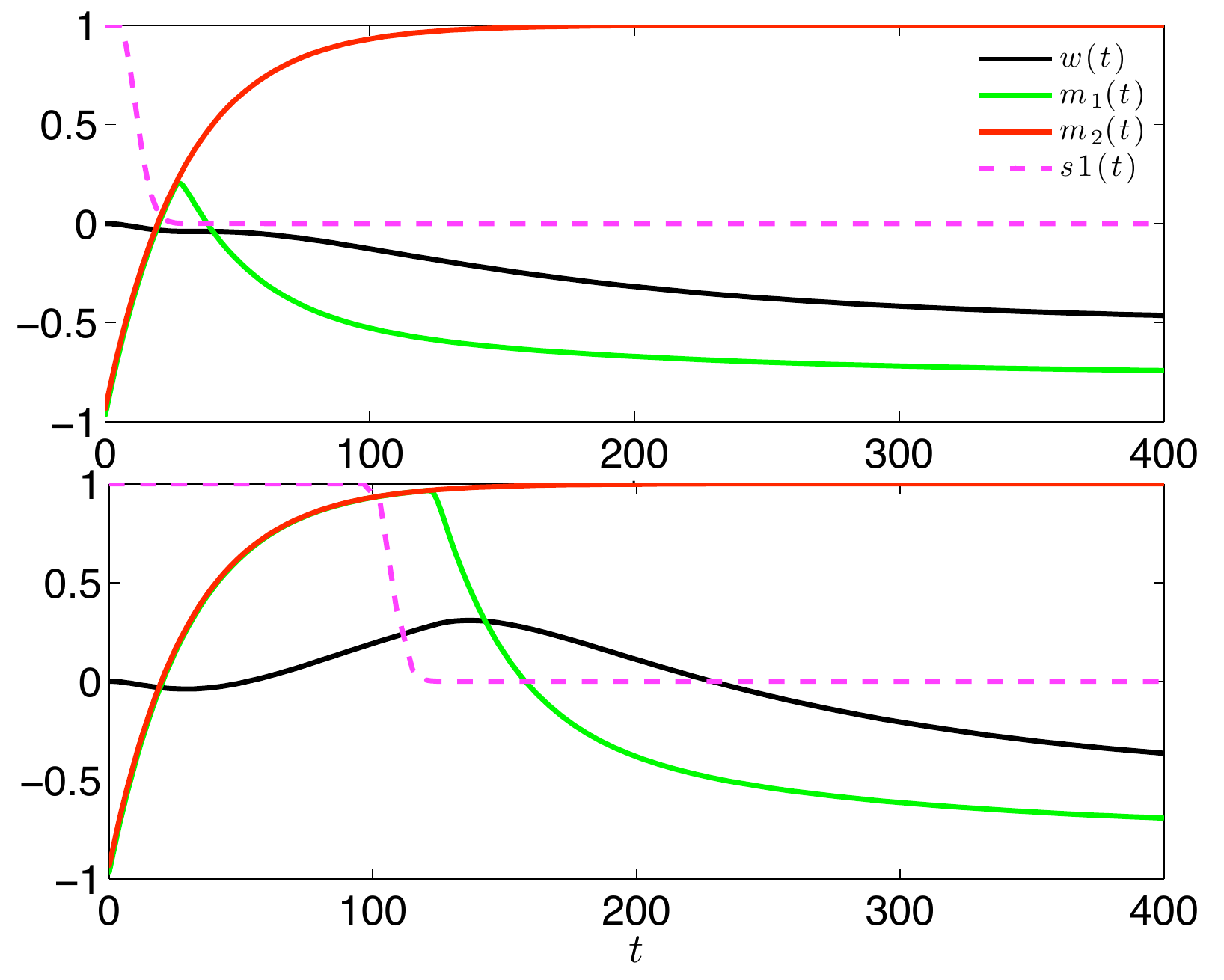}
\caption{\label{fig:highT}(Color on line)  Averages $w$, $m_1$ $m_2$ and signal $s_1$ (see the legend) as a function of time for Patient 1 (top panel) and Patient 2 (bottom panel). The level of noise is low and fixed at $\beta=0.8$.}
\end{center}
\end{figure}

\smallskip
Finally, when the level of noise is high, we expect that the effects due to interaction get more and more negligible. Indeed, in this model "low" and "high" levels of noise are not referred to a
specific ``critical value'', as this model does not break ergodicity by itself; conversely, since the noise level $\beta$ is coupled to the averaged
energy in the system $E = \langle J \sigma_1 \sigma_2 \rangle$, their product defines the levels, i.e., either $\beta E > 1$ , or $\beta E < 1$.
For instance, the case $\beta =0.8$ is shown in Fig.~\ref{fig:highT}: notice that independently of the duration of $s_1$, both the averages $m_1$ and $m_2$ relax to small values.

\smallskip

To summarize, according to the duration of the AIM  phase, which
is in turn  related to the success of the EBV strategy
(Sec.~\ref{sec:EBV}), we can get two possible scenarios. If the AIM
phase is rather fast, when the viral load has vanished, $B$ cells
are continuously activated, while the concentration of $T_K$ cells
recovers normal values hence we reach an equilibrium state
corresponding to an healthy carrier state.
\newline
Conversely, if AIM phase is prolonged, a strong correlation
between the active $B$ and $T_K$ lymphocytes can be accomplished;
as a result, when the viral load has vanished, again $B$ cells are
continuously activated but $T_K$ lymphocytes can maintain high
concentrations. This is the actualization of a conditional
reflex and, from a mathematically point of view it arises from the
large correlation $w$ stored, able to sustain the active status of
$T_K$.

\section{Conclusion and Discussion}
The Chronic Fatigue Syndrome (CFS) has
been studied for almost thirty years in the whole biological, medical and psychological world, being identified with tens of medical terms; as well, a full consensus on its genesis, physiopathology and treatment has not been reached yet.

It is just the lack of a clear-cut picture that makes theoretical
models very useful tools in order to get information about the
possible causes of this disease. The present work aims to take
advantage of the statistical-mechanics lenses to investigate why and
how the CFS may establish. Of course, the model cannot simulate
the whole immune system, which is much too complicated, rather, it
has to get a compromise between simplification and inclusion of
most important characters, the latter chosen according to
experimental facts.

Our framework is inspired by the well-known conditional  reflex
phenomenon in neurobiology, which, from a statistical mechanics perspective, can be recovered in terms of
thermodynamic relaxation of complex systems \cite{guerra}.
Given two agents (e.g. two neurons or $B$ and $T_K$ lymphocytes) and a coupling
(e.g. the synapse or the $T_H$ lymphocytes) joining them, provided that two agents are
contemporary stimulated by two related signals (e.g. neural stimuli or antigens), then, even though one
of the signal is switched off, the two agents may remain both active; this realizes the so-called "associative learning".
\newline
More precisely, along this analogy, we find that, if the Epstein-Barr virus infection is
prolonged in time, $T_K$ and B cells can  reach high values of
concentration, moreover, the latter have time enough to produce
the image ebv**. As a result, when the (real) viral load has
vanished, B cells are still stimulated, while $T_K$ may maintain high
concentration levels since they have learnt the correlation
between presence-of-antigen and B-activation. Interestingly, this
learning process also implies that $T_K$ can get activated by-passing the
signal from $T_H$ cells and this has been recently found
experimentally.
\newline
Therefore, our model and our analysis suggest that the  CFS, meant
as a chronically active immune state, can arise from an
associative-learning phenomenon.
\newline
As a consequence, our results clarify why some individuals are affected by this disease while some
other are not, why some infections may drive it (we investigated the case of EBV) while some other may not:
Within our framework, the genesis of this chronically active state is guided by the ability of a particular virus (e.g. EBV)
to elude immuno-surveillance for a larger timescale with respect to the standard ones set by the other antigens. Clearly, as EBV-infected
patients may solve the infection in a wide range of times, the ones with the infection lasting for too long are natural
candidates to develop CFS.  Likewise, other pathogens, such as Citomegalovirus, can drive a long-term infection, hence allowing correlations
and non-trivial outcomes.
\newline
Furthermore, our suggestion may be exploited in future experiments in order to shed light on the
etiology of this syndrome; indeed, at least at a
theoretical physics level, unlearning processes are actually possible.

Finally, it is worth stressing the "guide role" of statistical
mechanics when modeling biological systems: in fact this approach,
constraining the system to respect thermodynamics, can provide a
working picture possibly inspiring experimental paths.
In this sense, relying on natural and minimal assumptions, we
evidence the emergence of a subtle
role by $T_K$ (able to bypass a standard double signal activation
provided by antigen plus helpers), and, indeed, we found its
consistency with experimental data. In fact in \cite{111}, it has been
documented that a CD8+ T receptors, that normally recognize MHC-I signals can exhibit dual specificity recognizing also an
antigen in the context of the MHC-II.

\section{Appendix: Evolution toward steady states of the system}

The system, whose dynamics is investigated Monte Carlothrough the paper, can be
resumed as follows:
\newline
$\sigma_1$ and $\sigma_2$ are the effector cells (i.e., neurons in the neurobiology counterpart), while 
$J_{12}$ is the helper cell (synapse); moreover we name $s_1,s_2$  the antigens (external fields)
acting respectively on $\sigma_1,\sigma_2$.
Lymphocytes can be either active or quiescent ($\sigma_i = \pm1$) and can be grouped into clones. Each clone is made up of a huge amount cells expressing the same idiotipicity and it is expected to be homogeneous. Analogously, in statistical mechanics, we can apply the ergodic hypothesis, and move from their time-average to their ensemble average, using equivalently
\begin{eqnarray}
m_i &= &\int_0^T \sigma_i(t)dt = \frac1N \sum_{j=1}^N \sigma_i^{(j)},\\
w &=& \int_0^T J (t)dt = \frac1N \sum_{j=1}^N J^{(j)},
\end{eqnarray}
where $i=1,2$ and the index $(j)$ denotes the element within the clone considered.
We also stress that both the support $T$ of the integral and the size $N$ of the clone are large: In the limit of
$T,N \to \infty$ the variable $m \in \mathbb{R}$ hence allowing a continuous description in terms of Langevin equations.
\newline
Each of the variable experiences the same structure of
(external/internal) fields, namely (calling $\sigma_{3} = J_{12}$ to preserve the symmetry)
\begin{equation} \langle
\sigma_i \rangle = \langle \tanh\Big( \beta \sigma_{i+1}\sigma_{i+2} + s_i
\Big) \rangle = a_i + b_i \langle \sigma_{i+1}\sigma_{i+2} \rangle,\end{equation} where
\begin{eqnarray}
a_i &=& \frac12 \left[ \tanh(\beta+s_i)+ \tanh(-\beta+s_i) \right], \\
b_i &=& \frac12 \left[ \tanh(\beta+s_i)- \tanh(-\beta+s_i) \right].
\end{eqnarray}
Overall, we have the following system
\begin{eqnarray} \label{conti1}
\langle \sigma_1 \rangle &=& \langle \tanh(\beta J \sigma_2 + s_1) \rangle =  \tanh(\beta J \langle\sigma_2\rangle + s_1), \\
\langle J \rangle &=& \langle \tanh(\beta \sigma_1\sigma_2) \rangle=
\langle \sigma_1\sigma_2 \rangle \tanh(\beta), \\ \label{conti2}
\langle \sigma_2 \rangle &=& \langle \tanh(\beta J \sigma_1 + s_2) \rangle =  \tanh(\beta J \langle\sigma_1\rangle + s_1,
\end{eqnarray}
where in the last passage of eq.s (\ref{conti1},\ref{conti2}) we assumed the mean field approximation, namely
$$\langle f( m ) \rangle \sim  f (\langle m  \rangle),$$ as we are interested in the average behavior.
\newline
The time scale of $J$ is $\Theta \ll \tau$,
being $\tau$ the time scale of the $\sigma$ and we pose
$\frac{1}{\Theta}+\frac{1}{\tau}=\frac{1}{\tau'}$,
$\frac{2}{\tau}+\frac{1}{\Theta}=\frac{1}{\tau''}$.
For the sake of clearness, while we used the milder notation with $m,w$ in the text to lighten the notation, here we develop the whole theory
using the brackets. The averages are assumed to evolve according to
\begin{eqnarray}
\tau \frac{d\langle \sigma_1 \rangle}{dt} &=& -\langle \sigma_1
\rangle + a_1 + b_1\langle J \sigma_2 \rangle, \\
\Theta \frac{d\langle J \rangle}{dt} &=& - \langle J \rangle +
\tanh(\beta)\langle \sigma_1\sigma_2 \rangle, \\
\tau \frac{d\langle \sigma_2 \rangle}{dt} &=& -\langle \sigma_2
\rangle + a_2 + b_2\langle J \sigma_1 \rangle,
\end{eqnarray}
while the correlations evolve according to
\begin{eqnarray}
\frac{d\langle \sigma_1\sigma_2 \rangle}{dt} &=&
\frac{-2}{\tau}\langle \sigma_1\sigma_2\rangle + \frac{1}{\Theta}\Big(
a_1 \langle \sigma_2 \rangle + b_1 \langle J \rangle + a_2 \langle
\sigma_1 \rangle + b_2 \langle J \rangle \Big), \\
\frac{d\langle J\sigma_1 \rangle}{dt} &=&
\frac{-1}{\tau'}\langle J\sigma_1\rangle +
\frac{1}{\Theta}\tanh(\beta)\langle \sigma_2 \rangle +
\frac{1}{\tau}\Big( a_1 \langle J \rangle + b_1 \langle \sigma_2
\rangle \Big), \\
\frac{d\langle J\sigma_2 \rangle}{dt} &=&
\frac{-1}{\tau'}\langle J\sigma_2\rangle +
\frac{1}{\Theta}\tanh(\beta)\langle \sigma_1 \rangle +
\frac{1}{\tau}\Big( a_2 \langle J \rangle + b_2 \langle \sigma_1
\rangle \Big), \\
\frac{d \langle J \sigma_1 \sigma_2 \rangle}{dt} &=&
\frac{-2}{\tau''}\langle J \sigma_1 \sigma_2 \rangle + \frac{1}{\Theta}
\tanh(\beta) + \frac{1}{\tau}\Big( a_1 \langle J \sigma_2 \rangle
+ b_1 + a_2 \langle J \sigma_1 \rangle + b_2 \Big),
\end{eqnarray}
as in standard off-equilibrium statistical mechanics.
\newline
The source of $\langle \sigma_1 \rangle$ and of $\langle \sigma_2 \rangle$ are $s_1$ and $s_2$, respectively, while the source for $\langle
J \rangle$ is the internal correlation $\langle \sigma_1 \sigma_2
\rangle$.
\newline
According to the mutual value of $s_1$ and $s_2$, such a system experiences different regimes, whose dynamics
we are going to analyze.

\subsection{Regime $s_1=0, s_2=0$: No signalling.}

Langevin dynamics reduces to
\begin{eqnarray}
\Theta\frac{d\langle J \rangle}{dt} &=& - \langle J \rangle +
\tanh(\beta) \langle \sigma_1\sigma_2\rangle, \\
\tau\frac{d \langle \sigma_1 \rangle}{dt} &=& - \langle \sigma_1
\rangle + \tanh(\beta) \langle J \sigma_2 \rangle, \\
\tau\frac{d \langle \sigma_2 \rangle}{dt} &=& - \langle \sigma_2
\rangle + \tanh(\beta) \langle J \sigma_1 \rangle, \\
\frac{d \langle \sigma_1\sigma_2 \rangle}{dt} &=& \frac{1}{\tau}\Big( -2 \langle
\sigma_1\sigma_2 \rangle + 2\tanh(\beta)\langle J \rangle \Big),\\
\frac{d\langle J \sigma_1 \rangle}{dt} &=& -\frac{1}{\tau'}\langle
J \sigma_1\rangle + \frac{1}{\tau'}\tanh(\beta)\langle \sigma_2
\rangle,\\
\frac{d\langle J \sigma_2 \rangle}{dt} &=& -\frac{1}{\tau'}\langle
J \sigma_2\rangle + \frac{1}{\tau'}\tanh(\beta)\langle \sigma_1
\rangle,\\
\frac{d \langle J \sigma_1 \sigma_2 \rangle}{dt} &=&
-\frac{1}{\tau''}\langle J \sigma_1\sigma_2\rangle +
\frac{1}{\tau''}\tanh(\beta).
\end{eqnarray}
As it is immediate to see, the global dynamics spreads over four
different independent sub-dynamics, namely $\langle \sigma_1
\rangle \Longleftrightarrow \langle J \sigma_2 \rangle$, $\langle
\sigma_2 \rangle \Longleftrightarrow \langle J \sigma_1 \rangle$,
$\langle \sigma_1\sigma_2 \rangle \Longleftrightarrow \langle J
\rangle$, $\langle J\sigma_1\sigma_2 \rangle$, whose asymptotic
regime is given by $\langle J \rangle = \langle \sigma_1 \sigma_2
\rangle = 0$, $\langle \sigma_1 \rangle = \langle J \sigma_2
\rangle = 0$, $\langle \sigma_2 \rangle = \langle J
\sigma_1\rangle = 0$, $\langle J \sigma_1 \sigma_2 \rangle=
\tanh(\beta)$.
\newline
The general solution of the problem can be obtained coupling the
four different sub-dynamics.
\newline
By the first set we get
\begin{eqnarray}
\frac{d\langle J \rangle}{dt} &=& -\frac{1}{\Theta}\langle J \rangle +
\frac{\tanh(\beta)}{\Theta}\langle \sigma_1 \sigma_2\rangle \\
\frac{d \langle \sigma_1 \sigma_2 \rangle}{dt} &=&
-\frac{2}{\tau}\langle \sigma_1\sigma_2\rangle +
\frac{2\tanh(\beta)}{\tau}\langle J \rangle.
\end{eqnarray}
whose matrix can be written as
$$\left (
\begin{array} {c c }
1/\Theta & -\tanh(\beta)/\Theta \\
-2\tanh(\beta)/\Theta & 2/\Theta \\
\end{array}
\right ).
$$
We can diagonalize the sub-dynamics by looking for solutions as
linear combinations as
\begin{equation}
Y(t)=a\langle J(t) \rangle + b \langle
\sigma_1(t)\sigma_2(t)\rangle,
\end{equation}
and we can associate to this new variable a characteristic
timescale $\bar{\tau}$ as
\begin{eqnarray}
&& \bar{\tau} \frac{ dY}{dt} = - Y \Longrightarrow \frac{a\bar{
\tau}}{\Theta}\Big( -\langle J \rangle + \tanh(\beta) \langle
\sigma_1\sigma_2\rangle \Big)\\
\nonumber
&+&\frac{2b\bar{\tau}}{\tau}\Big(
-\langle \sigma_1\sigma_2\rangle + \tanh(\beta)\langle J \rangle
\Big)= -a\langle J\rangle - b \langle \sigma_1\sigma_2 \rangle.
\end{eqnarray}
Namely we get the system
\begin{eqnarray}
\Big(1 - \frac{\bar{\tau}}{\Theta}
\Big)+2\frac{\bar{\tau}}{\tau}\tanh(\beta)b&=&0 \\
\frac{\bar{\tau}}{\Theta}\tanh(\beta) a + \Big( 1-2\frac{\bar{\tau}}{\Theta}
\Big)b&=&0.
\end{eqnarray}
Let us work out $\bar{\tau}(\beta)$:
\begin{equation}\label{delta}
\left ( \frac{1}{\bar{\tau}} \right)^2- \left(\frac{1}{\bar{\tau}}\right)\left[ \frac{1}{\Theta}
\frac{2}{\tau} \right] + \frac{2}{\tau \Theta}\frac{1}{\cosh^2(\beta)}=0,
\end{equation}
whose roots are
\begin{equation}\label{roots}
\frac{1}{\bar{\tau}_{1,2}\left(\beta\right)}=\frac12\left[ \frac{1}{\Theta} +
\frac{2}{\tau} \pm \sqrt{\left( \frac{1}{\Theta} + \frac{2}{\tau}
\right)^2 - \frac{8}{\tau \Theta \cosh^2(\beta)}}\right]. \end{equation}
Now we have to
solve for $a,b$ in $ Y = a \langle J \rangle + b \langle
\sigma_1 \sigma_2 \rangle$. We can define $a = 2c
\frac{\bar{\tau}}{\Theta}\tanh(\beta)$, $b = -c (1-
\frac{\bar{\tau}}{\Theta})$, by which
\begin{eqnarray}
Y_0(t) &=& c_0 \Big(2 \frac{\bar{\tau}_1}{\tau}\tanh(\beta)
\langle J \rangle - (1 - \frac{\bar{\tau}_1}{\Theta})\langle \sigma_1
\sigma_2 \rangle \Big) \\
Y_{12}(t) &=& c_{12} \Big( 2
\frac{\bar{\tau}_2}{\tau}\tanh(\beta)\langle J \rangle -
(1-\frac{\bar{\tau}_2}{\Theta})\langle \sigma_1 \sigma_2 \rangle \Big),
\end{eqnarray}
on which we can fix $c_0, c_{12}$ as $c_0 = \tau/
2\bar{\tau}_1\tanh(\beta)$, $c_{12}= -1/(1-\bar{\tau}_2/\Theta)$. From
eq.(\ref{delta}) we get $$\tanh(\beta) = \sqrt{\frac{\tau
\Theta}{2}}\sqrt{\left(\frac{1}{\bar{\tau}}-\frac{1}{\Theta}\right)\left(\frac{1}{\bar{\tau}}-\frac{2}{\tau}\right)}$$
and we can solve for $Y_0,Y_{12}$:
\begin{eqnarray}
Y_0(t) &=& \langle J \rangle +
\sqrt{\frac{\tau}{2\Theta}}\sqrt{\frac{\frac{1}{\Theta}-\frac{1}{\bar{\tau}_1}}{\frac{2}{\tau}-\frac{1}{\bar{\tau}_1}}}
\langle \sigma_1 \sigma_2\rangle, \\
Y_{12}(t) &=& \langle \sigma_1 \sigma_2 \rangle -
\sqrt{\frac{2\Theta}{\tau}}\sqrt{\frac{\frac{1}{\bar{\tau}_2}-\frac{2}{\tau}}{\frac{1}{\bar{\tau}_2}-\frac{1}{\Theta}}}
\langle J \rangle,
\end{eqnarray}
so we get the form
\begin{eqnarray}
Y_0(t)&=& \langle J(t) \rangle + A_0 \langle \sigma_1(t)
\sigma_2(t) \rangle,\\
Y_{12}(t) &=&\langle \sigma_1 (t) \sigma_2(t) \rangle - A_{12}
\langle J(t) \rangle,
\end{eqnarray}
with
$$
A_0=
\sqrt{\frac{\tau}{2\Theta}}\sqrt{\frac{\frac{1}{\Theta}-\frac{1}{\bar{\tau}_1}}{\frac{2}{\tau}-\frac{1}{\bar{\tau}_1}}},
\ \ \ A_{12} =
\sqrt{\frac{2\Theta}{\tau}}\sqrt{\frac{\frac{1}{\bar{\tau}_2}-\frac{2}{\tau}}{\frac{1}{\bar{\tau}_2}-\frac{1}{\Theta}}}.
$$
By the eigenvalues found in eq.s(\ref{roots}) we can build the
eigenvectors
$V_1=(V_{11},V_{12}),V_2=(V_{21},V_{22})$ as
$$
V_{11}=\tanh(\beta)/\Theta, \ V_{12} = \left ( \frac{1}{2T}-\frac{1}{\tau} -
\frac{1}{2}\sqrt{\Delta} \right),$$
$$
V_{21} = \tanh(\beta)/\Theta, \ V_{22} =
\left ( \frac{1}{2\Theta}-\frac{1}{\tau} + \frac{1}{2}\sqrt{\Delta} \right),
$$
being $\Delta = (1/\Theta + 2/\tau)^2 - \frac{8}{\Theta \tau
\cosh^2(\beta)}$, by which, finally we get
\begin{eqnarray}
\langle J(t) \rangle &=& C_1 V_{11} e^{-\frac{t}{\bar{\tau}_1}} +
C_2 V_{12} e^{-\frac{t}{\bar{\tau}_2}}, \\
\langle \sigma_1(t)\sigma_2(t) \rangle &=& C1 V_{21}
e^{-\frac{t}{\bar{\tau}_1}}+C_2 V_{22}
e^{-\frac{t}{\bar{\tau}_2}}.
\end{eqnarray}
By the second set we get

\begin{eqnarray}
\tau \frac{d \langle \sigma_1 \rangle}{dt} &=& - \langle \sigma_1
\rangle + \tanh(\beta)\langle J \sigma_2 \rangle,\\
\tau' \frac{d \langle J \sigma_2 \rangle}{dt} &=& - \langle J \sigma_2
\rangle + \tanh(\beta)\langle \sigma_1 \rangle.
\end{eqnarray}
whose matrix can be written as
$$\left (
\begin{array} {c c }
1/\tau & -\tanh(\beta)/\tau \\
-\tanh(\beta)/\tau' & 1/\tau' \\
\end{array}
\right ).
$$
Again we can write a solution in the general form $Y(t) = a\langle
\sigma_1 \rangle + b \langle J \sigma_2 \rangle$ and label
$\bar{\tau}$ its characteristic timescale such that
$$
\bar{\tau}\frac{d Y}{dt} = -Y
$$
$$
\Longrightarrow
\frac{\bar{\tau}}{\tau}  a \Big( - \langle \sigma_1 \rangle +
\tanh(\beta) \langle J \sigma_2 \rangle \Big) +
\frac{\bar{\tau}}{\tau'}  b \Big( -\langle J \sigma_2 \rangle +
\tanh(\beta) \langle \sigma_1 \rangle \Big) = $$
$$
= -a \langle \sigma_1
\rangle + b \langle J \sigma_2 \rangle,$$ and write the system
\begin{eqnarray}
(1-\frac{\bar{\tau}}{\tau}) a + \frac{\bar{\tau}}{\tau'}\tanh(\beta) b &=& 0, \\
\frac{\bar{\tau}}{\tau}\tanh(\beta) a +
(1-\frac{\bar{\tau}}{\tau'}) b &=& 0.
\end{eqnarray}
Again we can find the eigenvalues $\bar{\tau}^{-1}_{1,2}(\beta)$
as
\begin{equation}
\frac{1}{\bar{\tau}_{1,2}(\beta)} = \frac12 \Big( \frac{1}{\tau} +
\frac{1}{\tau'}\pm
\sqrt{(\frac{1}{\tau}+\frac{1}{\tau'})^2-\frac{4}{\tau
\tau'}(1-\tanh^2(\beta))} \Big)
\end{equation}
and write the general solution in the form
\begin{eqnarray}
\langle \sigma_1(t) \rangle &=&
\frac{C_1\tanh(\beta)}{\tau}e^{-\frac{t}{\bar{\tau}_1}} -
\frac{C_2\tanh(\beta)}{\tau}e^{-\frac{t}{\bar{\tau}_2}}, \\
\langle J\sigma_2(t) \rangle &=& -
\frac{C_1}{2}(\frac{1}{\Theta}-\sqrt{\Delta})e^{-\frac{t}{\bar{\tau}_1}}-
\frac{C_2}{2}(\frac{1}{\Theta}-\sqrt{\Delta})e^{-\frac{t}{\bar{\tau}_2}},
\end{eqnarray}
$\Delta$ being $(1/\tau + 1/\tau')^2 -
(4/\tau\tau')(1-\tanh(\beta))$.

\bigskip

For the two other subsystems we can proceed exactly as we did so
far and verify that, even though on different timescales, all
 the observables (magnetizations and correlations) converge to
zero.

\subsection{Regime $s_1=0, s_2=\infty$: One infinite signal.}

Let us start with the following conditions: $s_1=0, s_2= \infty$,
then we have $a_1=0, b_1=\tanh(\beta), a_2=1, b_2=0$ and the
evolution of the system can be written as
\begin{eqnarray}
\Theta \frac{d \langle J \rangle}{dt} &=& - \langle J \rangle +
\tanh(\beta) \langle \sigma_1 \sigma_2 \rangle, \\
\tau \frac{d \langle \sigma_1 \rangle}{dt} &=& - \langle \sigma_1
\rangle + \tanh(\beta) \langle J \sigma_2 \rangle, \\
\tau \frac{d \langle \sigma_2 \rangle}{dt} &=& -\langle \sigma_2
\rangle + 1, \\
\tau \frac{d\langle \sigma_1 \sigma_2 \rangle}{dt} &=& - 2 \langle
\sigma_1 \sigma_2 \rangle + \tanh(\beta)\langle J \rangle  +
\langle \sigma_1 \rangle,\\
\frac{d \langle J \sigma_1 \rangle}{dt} &=&
-\frac{1}{\tau'}\langle J\sigma_1 \rangle + \frac{1}{\tau'}
\tanh(\beta) \langle \sigma_2 \rangle, \\
\frac{d \langle J \sigma_2 \rangle}{dt} &=& - \frac{1}{\tau'}
\langle J \sigma_2 \rangle + \frac{1}{\Theta}\tanh(\beta)\langle
\sigma_1\rangle + \frac{1}{\tau}\langle J \rangle,\\
\frac{d \langle J \sigma_1 \sigma_2 \rangle}{dt} &=&
-\frac{1}{\tau''}\langle J \sigma_1 \sigma_2 \rangle +
\frac{1}{\tau} \langle J \sigma_1 \rangle +
\frac{1}{\tau'}\tanh(\beta).
\end{eqnarray}
The presence of a signal in the second channel bridges the two
sets $(\langle \sigma_1(t) \rangle, \langle J \sigma_2(t)
\rangle)$ and $(\langle \sigma_1(t) \sigma_2(t) \rangle, \langle
J(t) \rangle)$ such that the latter, if the system has experienced
the field enough time to learn correlations, will assume high
values: a signal in the second channel may induce a
response even in the first channel.

\bigskip

Let us divide the system again by considering  the following, natural sub-dynamics
\begin{eqnarray}
\Theta \frac{d \langle J \rangle}{dt} &=& - \langle J \rangle +
\tanh(\beta) \langle \sigma_1 \sigma_2 \rangle,\\
\nonumber
\tau \frac{d \langle \sigma_1 \sigma_2 \rangle}{dt} &=& -2 \langle
\sigma_1 \sigma_2 \rangle + 2 \tanh(\beta) \langle J \rangle\\
&+&[\langle \sigma_1 \rangle - \tanh(\beta)\langle J(t) \rangle].
\end{eqnarray}
We observe that the term $[\langle \sigma_1 \rangle -
\tanh(\beta)\langle J(t) \rangle]$ is a novelty with respect to
the same equations in the regime $s_1=0,s_2=0$ and represents the
"unlearning source": if the conditional reflex (which we are going
to introduce) is not reinforced, it will tend to vanish.
\newline
The source of such a conditional reflex can be found by looking at
the other subsystem, namely
\begin{eqnarray}
\tau \frac{d \langle \sigma_1 \rangle}{dt} &=& - \langle \sigma_1
\rangle + \tanh(\beta) \langle J \sigma_2 \rangle, \\
\nonumber
\frac{d\langle J \sigma_2 \rangle}{dt} &=& -\frac{1}{\tau'} \langle
J \sigma_2 \rangle + \frac{1}{\tau'}\tanh(\beta) \langle \sigma_1
\rangle \\
&+& \frac{1}{\tau}[\langle J \rangle - \tanh(\beta) \langle
\sigma_1 \rangle].
\end{eqnarray}
In fact, the new term $[\langle J \rangle - \tanh(\beta) \langle
\sigma_1 \rangle]$, namely the source of the conditional reflex, is a
learning term as it couples the slow channel $(\langle J \rangle,
\langle \sigma_1\sigma_2 \rangle)$ with the fast one $(\langle
\sigma_1 \rangle, \langle J \sigma_2 \rangle)$.
\newline
We can now start studying the dynamics in this regime by
considering the sub-dynamics of ($\langle J \rangle$, $\langle
\sigma_1 \rangle$, $\langle \sigma_1 \sigma_2 \rangle$, $\langle J
\sigma_2 \rangle$), whose dynamical system reads off as
\begin{eqnarray}
\tau \frac{d \langle \sigma_1 \rangle}{dt} &=& -\langle \sigma_1
\rangle + \tanh(\beta) \langle J \sigma_2 \rangle,\\
\Theta \frac{d \langle J \rangle}{dt} &=& - \langle J \rangle +
\tanh(\beta) \langle \sigma_1 \sigma_2 \rangle, \\
\tau \frac{d \langle \sigma_1 \sigma_2 \rangle}{dt} &=& -2 \langle
\sigma_1 \sigma_2 \rangle + \tanh(\beta)\langle J \rangle +
\langle \sigma_1 \rangle,\\
\frac{d \langle J \sigma_2 \rangle}{dt} &=& -\frac{1}{\tau'}
\langle J \sigma_2 \rangle + \frac{1}{\tau}\langle J \rangle +
\frac{1}{\Theta}\tanh(\beta) \langle \sigma_1 \rangle.
\end{eqnarray}
The associated matrix  can be written as
$$\left (
\begin{array} {c c c c}
1/\tau & 0 & 0 & -\tanh(\beta)/\tau \\
0 & 1/\Theta &  -\tanh(\beta)/\Theta & 0 \\
-\frac{1}{\tau} & -\frac{1}{\tau}\tanh(\beta) & \frac{2}{\tau} & 0 \\
-\frac{1}{\Theta}\tanh(\beta) & -\frac{1}{\tau} & 0 & \frac{1}{\tau'}
\end{array}
\right ).
$$
We can diagonalize the dynamics and look for solutions as linear
combinations like $Y(t)=a \langle \sigma_1 \rangle + b \langle J
\rangle + c \langle \sigma_1 \sigma_2 \rangle + d \langle J
\sigma_2 \rangle$, associating to this variable its characteristic
timescale $\bar{\tau}\Leftrightarrow Y$, and proceed as for the
former regime.
\newline
Skipping all the calculations for the sake of brevity we report
only the solution
$$
\langle \sigma_1(t) \rangle = c_1 x_1
e^{-\frac{t}{\bar{\tau}_1}}+ c_2 x^{'}_1
e^{-\frac{t}{\bar{\tau}_2}} + c_3 x_1^{''}
e^{-\frac{t}{\bar{\tau}_3}}+ c_4 x_1^{'''}
e^{-\frac{t}{\bar{\tau}_4}},
$$
$$
\langle J(t) \rangle = c_1 x_2 e^{-\frac{t}{\bar{\tau}_1}}+c_2
x_2^{'} e^{-\frac{t}{\bar{\tau}_2}}+c_3 x_2^{''}
e^{-\frac{t}{\bar{\tau}_3}}+c_4 x_2^{'''}
e^{-\frac{t}{\bar{\tau}_4}},
$$
$$
\langle \sigma_1(t)\sigma_2(t) \rangle = c_1 x_3 e^{-\frac{t}{\bar{\tau}_1}} + c_2 x_3^{'}
e^{-\frac{t}{\bar{\tau}_2}} + c_3 x_3^{''}
e^{-\frac{t}{\bar{\tau}_3}} + c_4 x_3^{'''}
e^{-\frac{t}{\bar{\tau}_4}}, \
$$
$$
 \langle J(t) \sigma_2(t) \rangle
= c_1 x_4 e^{-\frac{t}{\bar{\tau}_1}}+c_2 x_4^{'}
e^{-\frac{t}{\bar{\tau}_2}} + c_3 x_4^{''}
e^{-\frac{t}{\bar{\tau}_3}}+c_4 x_4^{'''}
e^{-\frac{t}{\bar{\tau}_4}},
$$
all the $x$'s being the component of the following eigenvectors
$V_1=(x_1,x_2,x_3,x_4)$, $V_2=(x_1^{'},x_2^{'},x_3^{'},x_3^{'})$,
$V_3=(x_1^{''},x_2^{''},x_3^{''},x_3^{''})$,
$V_4=(x_1^{'''},x_2^{'''},x_3^{'''},x_3^{'''})$:
\begin{eqnarray}
\nonumber
x_1 &=& \frac{1}{\tau}\tanh(\beta),  \\
\nonumber
x_2 &=& \frac12 \Big( \frac{1}{\tau}-\frac{1}{\Theta} - \sqrt{\Delta}
\Big),  \\
\nonumber
x_3 &=& \frac{1}{\tau}\tanh(\beta), \\
\nonumber
x_4 &=& \frac12 \Big( \frac{1}{\tau}-\frac{1}{\Theta}-\sqrt{\Delta}
\Big),
\end{eqnarray}
\begin{eqnarray}
\nonumber
x^{'}_1 &=& \frac{1}{\tau}\tanh(\beta),  \\
\nonumber
x^{'}_2 &=& \frac12 \Big( \frac{1}{\tau}-\frac{1}{\Theta}
+\sqrt{\Delta}
\Big),  \\
\nonumber
x^{'}_3 &=& \frac{1}{\tau}\tanh(\beta), \\
\nonumber
x^{'}_4 &=& \frac12 \Big( \frac{1}{\tau}-\frac{1}{\Theta}+\sqrt{\Delta}
\Big),
\end{eqnarray}
\begin{eqnarray}
\nonumber
x^{''}_1 &=& \frac{1}{\tau}\tanh(\beta),  \\
\nonumber
x^{''}_2 &=& \frac12 \Big( \frac{1}{\tau}-\frac{1}{\Theta}
+\sqrt{\Delta}
\Big),  \\
\nonumber
x^{''}_3 &=& \frac{\Theta}{\tanh(\beta)}\Big[\frac{3}{2\tau \Theta}-\frac{1}{2\Theta^2}-\frac{1}{\tau^2}-\frac{1}{\tau \Theta}\tanh^2(\beta) -\sqrt{\Delta}(\frac{1}{\tau}-\frac{1}{2\Theta}) \Big], \\
\nonumber
x^{''}_4 &=& -\frac12 \Big(
\frac{1}{\tau}-\frac{1}{\Theta}+\sqrt{\Delta} \Big),
\end{eqnarray}
\begin{eqnarray}
\nonumber
x^{'''}_1 &=& \frac{1}{\tau}\tanh(\beta),  \\
\nonumber
x^{'''}_2 &=& \frac12 \Big( \frac{1}{\tau}-\frac{1}{\Theta}
-\sqrt{\Delta}
\Big),  \\
\nonumber
x^{'''}_3 &=& \frac{\Theta}{\tanh(\beta)}\Big[\frac{3}{2\tau \Theta}-\frac{1}{2\Theta^2}-\frac{1}{\tau^2}-\frac{1}{\tau \Theta}\tanh^2(\beta) -\sqrt{\Delta}(\frac{1}{\tau}-\frac{1}{2\Theta}) \Big], \\
\nonumber
x^{'''}_4 &=& -\frac12 \Big(
\frac{1}{\tau}-\frac{1}{\Theta}+\sqrt{\Delta}
\Big), \\
\end{eqnarray}
being $\Delta=1/\tau^2 + 1/\Theta^2 + 2(2\tanh^2(\beta)-1)/(\tau \Theta)$.
\newline
The remaining sub-dynamics of $(\langle \sigma_2 \rangle, \langle
J \sigma_1 \rangle, \langle J \sigma_1 \sigma_2 \rangle)$ is
depicted by the system
\begin{eqnarray}
\tau \frac{d \langle \sigma_2 \rangle}{dt} &=& -\langle \sigma_2
\rangle +1,\\
\tau' \frac{d \langle J \sigma_1 \rangle}{dt} &=& - \langle J
\sigma_1 \rangle + \langle \sigma_2 \rangle, \\
\frac{d \langle J \sigma_1 \sigma_2 \rangle}{dt} &=&
-\frac{1}{\tau''} \langle J \sigma_1 \sigma_2 \rangle +
\frac{1}{\tau} \langle J \sigma_1 \rangle +
\frac{1}{\tau'}\tanh(\beta).
\end{eqnarray}
By applying the framework previously shown several times, we obtain
the solutions
\begin{eqnarray}
\nonumber
&&\langle \sigma_2 \rangle = c_1 e^{-\frac{t}{\bar{\tau}_1}}+1,\\
\nonumber
&&\langle J \sigma_1 \rangle = c_1 \frac{\Theta}{\tau'}
e^{-\frac{t}{\bar{\tau}_1}}+c_2 e^{-\frac{t}{\bar{\tau}_2}}+1,\\
\nonumber
&&\langle J \sigma_1 \sigma_2 \rangle = c_1 \frac{\Theta}{\tau}
e^{-\frac{t}{\bar{\tau}_1}} + c_2 e^{-\frac{t}{\bar{\tau}_2}} +
c_3 e^{-\frac{t}{\bar{\tau}_3}}+
\tau^{''}\left[\frac{1}{\tau}+\frac{1}{\tau'}\tanh(\beta)\right].
\end{eqnarray}

\subsection{Regime $s_1=\infty,s_2=\infty$: Two infinite signals.}

Let us start with the following conditions: $s_1= \infty, s_2=
\infty$, then we have $a_1=a_2=1, b_1=b_2=0)$ and the evolution of
the system can be written as
\begin{eqnarray}
\nonumber
&&\Theta \frac{d \langle J \rangle}{dt} = - \langle J \rangle +
\tanh(\beta) \langle \sigma_1 \sigma_2 \rangle, \\
\nonumber
&&\tau \frac{d \langle \sigma_1 \rangle}{dt} = - \langle \sigma_1
\rangle + 1, \\
\nonumber
&&\tau \frac{d \langle \sigma_2 \rangle}{dt} = -\langle \sigma_2
\rangle + 1, \\
\nonumber
&&\tau \frac{d\langle \sigma_1 \sigma_2 \rangle}{dt} = - 2 \langle
\sigma_1 \sigma_2 \rangle + \langle \sigma_1 \rangle  +
\langle \sigma_2 \rangle,\\
\nonumber
&&\frac{d \langle J \sigma_1 \rangle}{dt} =
-\frac{1}{\tau'}\langle J\sigma_1 \rangle + \frac{1}{\Theta}
\tanh(\beta) \langle \sigma_2 \rangle + \frac{1}{\tau}\langle J \rangle, \\
\nonumber
&&\frac{d \langle J \sigma_2 \rangle}{dt} = - \frac{1}{\tau'}
\langle J \sigma_2 \rangle + \frac{1}{\Theta}\tanh(\beta)\langle
\sigma_1\rangle + \frac{1}{\tau}\langle J \rangle,\\
\nonumber
&&\frac{d \langle J \sigma_1 \sigma_2 \rangle}{dt} =
-\frac{1}{\tau''}\langle J \sigma_1 \sigma_2 \rangle +
\frac{1}{\Theta}\tanh(\beta) + \frac{1}{\tau}\langle J \sigma_2 \rangle
+ \frac{1}{\tau}\langle J \sigma_1 \rangle,
\end{eqnarray}
whose solutions, in complete analogy with the previously introduced
methodology, can be obtained as
\begin{eqnarray}
\nonumber
\langle \sigma_1(t)\rangle &=& c_2 e^{-\frac{t}{\tau}}+1,\\
\nonumber
\langle J(t)\rangle &=& c_1 e^{-\frac{t}{\tau}}+(c_2+c_3)\frac{\tanh(\beta)}{\Theta(\frac{1}{\Theta}-\frac{1}{\tau})}e^{-\frac{t}{\tau}}
+ c_4\frac{\tanh(\beta)}{\Theta(\frac{1}{\Theta}-\frac{2}{\tau})}e^{-\frac{2t}{\tau}}+\tanh(\beta) ,\\
\nonumber
\langle \sigma_2(t)\rangle &=& c_3 e^{-\frac{t}{\tau}}+1,\\
\nonumber
\langle \sigma_1 (t) \sigma_2 (t) \rangle &=&
(c_2+c_3)e^{-\frac{t}{\tau}}+c_4e^{-\frac{2t}{\tau}}+1,\\
\nonumber
\langle J(t) \sigma_1(t) \rangle &=&
c_1e^{-\frac{t}{\Theta}}+\frac{\tanh(\beta)}{(\frac{1}{\Theta}-\frac{1}{\tau})}(\frac{c_2}{\tau}+\frac{c_3}{\tau})e^{-\frac{t}{\tau}}+c_4\frac{\tanh(\beta)}{\tau
\Theta (\frac{1}{\Theta}-\frac{2}{\tau})(\frac{1}{\Theta}-\frac{1}{\tau})}e^{-\frac{2t}{\tau}}+c_5e^{-\frac{t}{\tau'}}+\tanh(\beta),\\
\nonumber
\langle J(t) \sigma_2(t) \rangle &=&
c_1e^{-\frac{t}{\Theta}}+\frac{\tanh(\beta)}{(\frac{1}{\Theta}-\frac{1}{\tau})}(\frac{c_2}{\tau}+\frac{c_3}{\tau})e^{-\frac{t}{\tau}}
+c_4\frac{\tanh(\beta)}{\tau \Theta
(\frac{1}{\Theta}-\frac{2}{\tau})(\frac{1}{\Theta}-\frac{1}{\tau})}e^{-\frac{2t}{\tau}}+c_6e^{-\frac{t}{\tau'}}+\tanh(\beta),\\
\nonumber
\langle J(t) \sigma_1(t) \sigma_2(t) \rangle &=& 2c_1
e^{-\frac{t}{\Theta}}+
\frac{\tanh(\beta)}{\tau(\frac{1}{\Theta}-\frac{1}{\tau})}(c_2+c_3)e^{-\frac{t}{\tau}}
+c_4\frac{\tanh(\beta)}{\tau^2
(\frac{1}{\Theta}-\frac{2}{\tau})(\frac{1}{\Theta}-\frac{1}{\tau})}e^{-\frac{2t}{\tau}}+ \\ 
\nonumber
&+& c_7e^{-\frac{t}{\tau''}}+\tanh(\beta)  + (c_5+c_6)e^{-\frac{t}{\tau'}} + \tanh(\beta).
\end{eqnarray}

\section*{Acknowledgements}

This work is supported by the Italian Ministry for Education and Research FIRB grant number RBFR08EKEV.
\newline
AB is partially funded by GNFM (Gruppo Nazionale per la Fisica
Matematica) which is also acknowledged.
\newline
FG and EA are partially funded by INFN (Istituto Nazionale di Fisica Nucleare) which is also acknowledged.
\newline
Furthermore we are grateful to Sapienza Universita' di Roma for its contribution to our research.



\end{document}